\documentclass[twocolumn,superscriptaddress,showpacs,preprintnumbers,showkeys,amsmath,amssymb,aps,floatfix]{revtex4-2}
\usepackage{array}[=2016-10-06]

\usepackage{graphicx} 
\usepackage{amsmath}
\usepackage{mathtools}
\usepackage{xcolor}
\usepackage{graphicx}
\usepackage{bm}
\usepackage[colorlinks=true, citecolor=blue, urlcolor=blue, linkcolor=blue]{hyperref}
\usepackage[normalem]{ulem}
\usepackage{soul}
\usepackage{siunitx}
\usepackage{multirow}
\usepackage[version=4]{mhchem}
\usepackage{enumitem}

\usepackage{diagbox} 

\usepackage{comment}

\newcommand{\CCOC}{Ca$_2$CuO$_2$Cl$_2$}
\newcommand{\NaCCOC}{Na$_x$Ca$_{2-x}$CuO$_2$Cl$_2$}
\newcommand{\VCCOC}{Ca$_{2-x}$CuO$_2$Cl$_2$}

\newcommand{\LCO}{La$_{2}$CuO$_4$}

\newcommand{\BSCCO}{Bi$_{2}$Sr$_2$Ca$_{n-1}$Cu$_n$O$_{2n+4+x}$}

\newcommand{\LaTOF}{La$_{2-x}$M$_x$CuO$_4$ (M=Ba,Sr)}
\newcommand{\YBCO}{YBa$_2$Cu$_3$O$_{7-x}$}

\newcommand{\IR}{Infra-Red}

\begin{document}

\title{Raman scattering in cuprate oxychlorides high-temperature superconductors single-crystals}

\author{Chafic Fawaz}
\email{chafic.fawaz@kit.edu}
\altaffiliation{Current address: Institute for Quantum Materials and Technologies, Karlsruhe Institute of Technology, D-76021 Karlsruhe, Germany}
\author{Yingzheng Gao}
\author{Luca Laveder}
\author{Lorenzo Menon}
\author{Owen Moulding}
\affiliation{Institut NEEL CNRS/UGA UPR2940 - 25 rue des Martyrs BP 166 - 38042 Grenoble cedex 9}

\author{Rolf Heid}
\affiliation{Institute for Quantum Materials and Technologies, Karlsruhe Institute of Technology, D-76021 Karlsruhe, Germany}

\author{Blair W. Lebert}
\affiliation{IMPMC-Sorbonne Universit\'es, Universit\'e Pierre et Marie Curie, CNRS, IRD, MNHN 4, place Jussieu, 75252 Paris, France}
\affiliation{Synchrotron SOLEIL, L'Orme des Merisiers, Saint-Aubin, 91192 Gif-sur-Yvette Cedex, France}

\author{Christophe Bellin}
\author{Keevin B\'eneut}
\affiliation{IMPMC-Sorbonne Universit\'es, Universit\'e Pierre et Marie Curie, CNRS, IRD, MNHN 4, place Jussieu, 75252 Paris, France}

\author{David Santos-Cottin}
\affiliation{Department of Physics, University of Fribourg, 1700 Fribourg, Switzerland}

\author{Ikuya Yamada}
\affiliation{Department of Materials Science, Graduate School of Engineering, Osaka Metropolitan University, 1-1 Gakuen-cho, Naka-ku, Sakai, Osaka 599-8531, Japan}

\author{Yuichi Okazaki}
\affiliation{Department of Materials Science, Graduate School of Engineering, Osaka Metropolitan University, 1-1 Gakuen-cho, Naka-ku, Sakai, Osaka 599-8531, Japan}

\author{Hajime Yamamoto}
\altaffiliation[Permanent address: ]{Institute of Multidisciplinary Research for Advanced Materials (IMRAM), Tohoku Univ. Katahira 2-1-1, Aoba-ku, Sendai 980-8577, Japan}
\affiliation{Laboratory for Materials and Structures, Tokyo Institute of Technology, 4259 Nagatsuta, Midori-ku, Yokohama, 226-8503, Japan}

\author{Masaki Azuma}
\affiliation{Materials and Structures Laboratory, Institute of Integrated Research, Institute of Science Tokyo, 4259 Nagatsuta, Midori-ku, Yokohama 226-8503, Japan}

\author{Marie-Aude Measson} 
\email{marie-aude.measson@neel.cnrs.fr}
\author{Matteo d'Astuto}
\email{matteo.dastuto@neel.cnrs.fr}
\affiliation{Institut NEEL CNRS/UGA UPR2940 - 25 rue des Martyrs BP 166 - 38042 Grenoble cedex 9}

\date{\today}

\begin{abstract}
We investigate the Raman response of sodium-doped cuprate oxychloride \NaCCOC\ high-temperature superconductors across their entire phase diagram, from the antiferromagnetic to the superconducting region.
In addition to the expected Raman-active phonon modes, we detect several additional modes that can be interpreted as being excited via a resonance process enabled by strong electron--phonon coupling. To verify the resonance effect, the Raman response at different incident photon energies was measured.
At high energies, there is a well-defined $B_{1g}$ mode in the antiferromagnetic phase, which can be interpreted as a bimagnon
We follow its temperature and doping dependence, providing information on the multimagnon excitation in these cuprates, which can be theoretically linked to the exchange interaction.
\end{abstract}


\maketitle

\section{Introduction}
\begin{figure}[htb]
\includegraphics[width=0.45\linewidth]{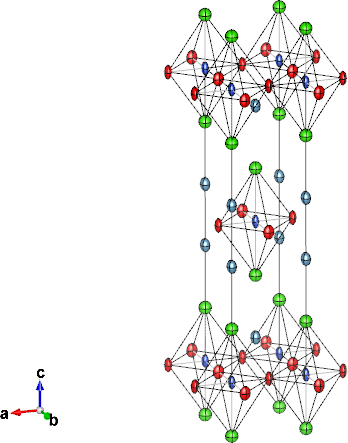}
\caption{\label{ccoc structure} (Color online) The crystallographic unit cell of \CCOC{} with Ca as cyan, Cu as blue, O as red, and Cl as green. The square coordination of Cu with its four nearest-neighbor O ions in the CuO$_2$ planes is shown. The Cl ions are located at the apical sites below/above the Cu ions. The atomic coordinates and displacement ellipsoids are from single-crystal diffraction detailed in Ref.~\onlinecite{Baptiste:wm4087}.}
\end{figure}

Raman scattering has been extensively used to investigate the vibrational, magnetic, and electronic properties of most high-temperature superconducting cuprates \cite{Devereaux_2007}.
The oxychloride cuprate \NaCCOC\ studied here is, to our knowledge, the only exception to date, despite being a particularly interesting member of the cuprate family \cite{Hiroi1994,Kohsaka2002,Bacq-Labreuil2025}.
These compounds are strongly two-dimensional, like the BSCCO family (\BSCCO), yet they crystallize in the simple I4/mmm structure of the La214 family (\LaTOF) in the high-temperature tetragonal (HTT) phase \cite{Fujita_1987,PhysRevB.35.7191,PhysRevB.83.104506}.
The crystal structure of M$_x$Ca$_{2-x}$CuO$_2$Cl$_2$ (with M = Na, K, or a Ca vacancy), shown in Fig.~\ref{ccoc structure}, can be obtained from the HTT structure by replacing the apical oxygen with chlorine.
Consequently, unlike La214 and \YBCO\ (Y123), these materials exhibit no known structural transitions with temperature or doping and show no misfit between the CuO$_2$ planes and the charge-reservoir blocks, as in BSCCO.
This makes them a useful model system for comparison with realistic implementations of the Hubbard model \cite{Bacq-Labreuil2025}.

In recent years, several studies have unveiled the phonon \cite{zenitani2005, d2013phonon, Lebert_2020, PhysRevResearch.4.033004} and magnetic excitations \cite{PhysRevB.95.155110,Lebert_2023} of \NaCCOC\ using both resonant and non-resonant inelastic X-ray scattering (RIXS and IXS). 
The dispersion of phonons in vacancy-doped (\VCCOC) and sodium-doped (\NaCCOC) has been measured using IXS \cite{d2013phonon,Lebert_2020,PhysRevResearch.4.033004} and calculated by Density Functional Theory (DFT) \cite{Lebert_2020}.
Using Infra-Red (IR) spectroscopy, Zenitani \textit{et al.} \cite{zenitani2005} have measured spectra ranging from 31 to 496 meV. They observed three absorption-peak dips, two corresponding to $E_u$ and one to an undefined mode. Patterson \cite{Patterson2008} has calculated the phonon energies in \CCOC\ using Hybrid Density Functional Theory (hybrid-DFT) at the zone-center $\Gamma$ and at the zone boundary for some modes.

In this work, using Raman spectroscopy with different laser wavelengths, we measured four different dopings, covering the entire phase diagram of the oxychloride cuprate, most notably:
\begin{enumerate}[noitemsep, topsep=0pt]
  \item Undoped, antiferromagnetic parent compound \CCOC{}, with T$_N\sim$250 K \cite{Vaknin1997}.
  \item Underdoped, non-superconducting sample, for $x=0.06$ (UD1).
  \item Underdoped, superconducting sample, with T$_c$ = 13 K, for $x=0.10$ (UD2).
  \item Maximally doped, superconducting sample, with T$_c$ = 27 K, for $x=0.18$ (OP).
\end{enumerate}

In addition to the expected phonon modes that are Raman active, we detected several modes that can be interpreted as single and double-phonons and their harmonics, which are excited $via$ a resonance process thanks to strong electron-phonon coupling, as already reported for other cuprates \cite{Sugai1989,Reedyk_1994,popovic2001,sugai2003,sugai2004}.
 
Finally, we detect a well-defined, high-energy $B_{1g}$ mode in the antiferromagnetic sample, which can be interpreted as a bimagnon, as has been reported previously for other cuprates \cite{Lyons1988,PhysRevB.42.1045,Sugai1988}.
We follow its temperature and doping dependence, giving valuable information on the multimagnon excitation in these cuprates, allowing us to better understand their contribution to the magnetic dynamics as seen, \textit{e.g}, by RIXS~\cite{Lebert_2023}

\section{Methods}

\subsection{Crystal growth and characterization}

Single crystals of the undoped, antiferromagnetic parent compound \CCOC{} were grown by flux, as described in Ref. [\onlinecite{Baptiste:wm4087}]. 
Na-doped single-crystals with x=0.06 and x=0.10 were synthesized using the flux method under a high-pressure/high-temperature as described in Ref. [\onlinecite{Bacq-Labreuil2025}] for x=0.06 and Ref. [\onlinecite{Lebert_2023}] for x=0.10. Additional details for the synthesis of the sample with x=0.06 will be given elsewhere \cite{NCCOCsyntNEEL}.

Finally, the x=0.18 Na-doped sample was also synthesized using the same method. The Pt capsule was filled with the mixture of \CCOC{}, NaClO$_4$ (flux, oxidizer, and Na source), and NaCl (flux, and Na source), and then put into the (Mg,Ca)O pressure-transmitting medium with a graphite furnace.  The medium was compressed to 8 GPa using a Walker-type high-pressure apparatus available at the Department of Materials Science of the Osaka Metropolitan University. The sample was heated to 1523 K in 30 min, held at this temperature for 30 min, cooled to 1073 K over 480 min (8 hours), and quenched to room temperature. After cooling, the pressure was released.

To determine the doping of the superconducting samples, we measured their onset critical temperature using an MPMS3 Quantum Design\copyright~ SQUID magnetometer.

The crystalline quality of the samples was checked, and the orientation of the facets was determined, using two 4-circle diffractometers with k-geometry goniometers and using a Mo anode source: an Oxford Diffraction Xcalibur equipped with a Sapphire CCD detector for preliminary experiments on undoped samples, and a Bruker D8 Venture equipped with a Photon III detector for the main experiment on all dopings, which we describe below.  

\begin{figure*}[th]
\centering
\includegraphics[width=1.0\textwidth]{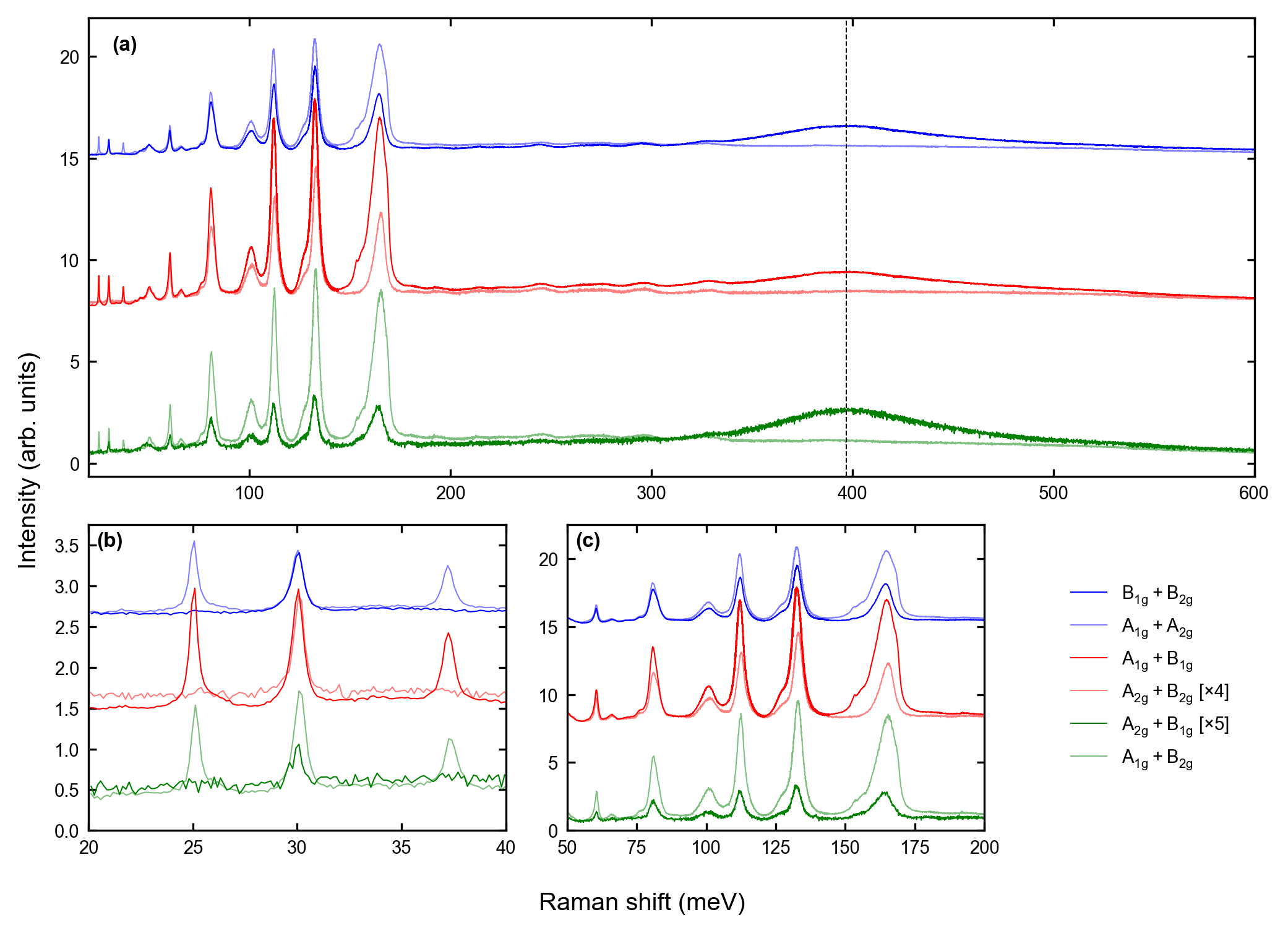}
\caption{(a) Raman spectroscopy measurements of antiferromagnetic \CCOC{} polarization in the back-scattering geometry measured at 30~K. Intensities are normalized to the area of the A$_{1g}$ mode at 30 meV. The spectra of a crossed/uncrossed pair are offset by a constant shift for clarity and plotted in the same color. The dashed gray vertical line indicates the peak energy of the B$_{1g}$ mode at around 400 meV.
(b) Details of the low (20-40 meV) and (c) intermediate (40-200 meV) energy regions from the top panel plot, with an adapted offset, and the same color code for polarization sets (see text for details).
}
\label{fig:CCOC_per_ramanpol}
\end{figure*}

\subsection{Raman}
We performed two series of experiments on two different Raman set-ups and cryostats. The first focused only on the undoped, antiferromagnetic parent compound \CCOC{}, but used back-scattering and both linear and circular polarizations for the incident and scattered light, allowing a complete determination of the mode's character. The instrument is located at the IMPMC (Sorbonne University, Paris), and in the following text, we refer to this first setup as the ``Paris-setup". The second set-up used only linear polarisation, with a $30^\circ$ reflection scattering angle, but spanned a wider sample doping and temperature range, included several light colors to study resonance effects, and allowed very high energy resolution and a low energy cutoff, close to the Rayleigh elastic scattering. This second instrument is located at the NEEL Institute (CNRS, Grenoble), and we label the first setup ``Grenoble-setup" hereafter. 

In more detail, the first investigation was conducted as a function of temperature over the 30-300 K range using a Jobin-Yvon/Horiba HR-460 spectrometer equipped with a monochromator with 1500 grooves/mm and a Peltier-cooled Andor CCD-2562. The Raman signal was excited using an Ar laser at $\lambda=514.5$ nm ($E_{incident} = 2.41$ eV), focused to a 2~$\mu$m spot, and collected in backscattering geometry. The set-up was optimized for high brilliance but relatively poor resolution, which was well suited to the very broad bimagnon signal. 

To perform this first set of experiments, we built and commissioned an adaptation for polarized Raman spectroscopy. The need for a larger energy region to capture a bimagnon meant stitching the spectra together, even with our low-groove-density grating. This was due to the fact that the spectrometer efficiency varied with the angle of the diffraction grating as we changed the energy range, since we kept a small region-of-interest (ROI) on the two-dimensional CCD detector to reduce background.
We optimized the background to study very broad energy signals, as is the case for the bimagnon one.  Note also that the gratings are sensitive to the incident polarization, acting as vertical polarizers. For this reason, we optimized the optical setup, always selecting linearly polarized vertical light to enter the spectrometer via a polarizer. For the (V,V) and (H,V) configurations, we modified the incident polarization using a prism. The incident polarization was first reflected off a beam splitter at 90$^{\circ}$ towards the sample, reflected off the sample in backscattering, and returned through the beam splitter in transmission towards the polarizer before the spectrometer. To achieve other polarizations, we would insert either a quarter-wave or a half-wave plate, appropriately aligned, between the cryostat holding the sample and the beam splitter. It can be shown that by switching the initial polarization between LH and LV with the prism and selecting LV polarization before the spectrometer entrance, we could probe the crossed and uncrossed configurations.

A second set of measurements was done on a triple-stage TriVista spectrometer from Teledyne Princeton Instruments. High-energy spectra ($>$ 10 meV) were taken using a 600 grooves/mm grating on the first stage of the spectrometer.

We have taken measurements over a wide temperature range, ranging from 2 K to 300 K, and using three lasers as probes: a blue laser with $\lambda=$ 488 nm, $E_{incident} = 2.54$ eV, a green one with $\lambda=532$ nm, $E_{incident} = \SI{2.23}{eV}$, and a red laser with $\lambda=660$ nm, $E_{incident} = 1.88$ eV. The laser power was set to 10 mW due to time constraints. This power level induces sample heating; therefore, the actual sample temperatures are higher than the nominal temperatures indicated in the figures.

In Appendix \ref{selrul}, we provide a summary of the Raman selection rules for the D$_{4h}$ point group as a reminder for the reader, in order to facilitate the interpretation of the results and the discussion. 

To enable reliable comparison of Raman signals acquired at different excitation wavelengths, we corrected the data (used in this comparison) for the wavelength-dependent grating response, as described in Appendix~\ref{WL_correction}.

\subsection{Density Functional Theory Phonon calculations}

Density Functional Theory (DFT) calculations of the phonon modes near the zone center $\Gamma$ for \CCOC{} are carried out using the linear response or density-functional perturbation theory implemented in the framework of the mixed-basis pseudopotential method \cite{Heid}, as in a previous work, Ref.!\cite{Lebert_2020}. The lattice structure of \CCOC{} was fully relaxed prior to the phonon calculations. We have used the stoichiometry of the undoped parent compounds.

\section{Results}\label{Results}

\subsection{Preliminary complete polarisation study in undoped \CCOC\ at low temperature}
Using the first experimental set-up (the ``Paris-setup"), we performed a preliminary study in an undoped \CCOC{} sample which was aligned such that linear horizontal polarization would be along the $a$-axis, and the incident photon beam along the $c$-axis. Figure~\ref{fig:CCOC_per_ramanpol} shows the data at 30~K with the crossed and uncrossed polarization for horizontal/vertical normal (\textit{i.e.} $\overline{c}(ab,ab)c$ and $\overline{c}(a,\overline{b})c$) and 45$^\circ$ (\textit{i.e.} $\overline{c}(a,a)c$ and $\overline{c}(ab,a\overline{b})c$) rotated, as well as circular polarization, over the whole energy range measured. 

The low-energy part of the spectra is zoomed in Fig.~\ref{fig:CCOC_per_ramanpol}(b), where we can single out two sharp modes with well-defined A$_{1g}$ character, at about 25 and 37 meV, with a symmetrical Lorentzian shape and a full-width-half-maximum (FWHM) of $\sim$ 0.4 to 0.6 meV. In between them, a third mode at about 30 meV shares a similar shape, width, and intensity, but not a unique character. We note that it is weaker only when the polarization allows an A$_{2g}$ contribution, so we can suppose an A$_{1g}$+B$_{1g}$+B$_{2g}$ character. 

The intermediate energy region, between 40 and 200 meV, is shown in Fig.~\ref{fig:CCOC_per_ramanpol}(c). In this region, several very intense modes appear, more than 10 times the low-energy A$_{1g}$ modes. They show asymmetric lineshapes with FWHM from about 4 to 8 meV, and no single character; most probably, again, with an A$_{1g}$+B$_{1g}$+B$_{2g}$ character. 
We note that the Debye energy of the phonon modes in \CCOC{} is around 85 meV \cite{Lebert_2020}, while the observed mode in this energy range appears from about 50 up to 170 meV, well above the single-phonon Debye energy. Several other weaker modes, without well-defined character, appear in the range from 200 up to about 350 meV.

Finally, in the higher-energy-range measurements, we observe a very broad but intense mode centered around 400 meV, with a FWHM of 80 meV and a well-defined B$_{1g}$ character. 

\subsection{Doping, temperature, and resonance study in $\mathbf{Na_xCa_{2-x}CuO_2Cl_2}$}\label{Result2}

In the second experiment, we investigate four different dopings, sampling all phases of the oxychloride cuprate, including the undoped antiferromagnetic compound \CCOC{}, for comparison with the preliminary experiment. As previously mentioned, the other dopings are:
\begin{enumerate}[start=2, noitemsep, topsep=0pt]
    \item Underdoped, non-superconducting sample, for $x=0.06$ (UD1).
    \item Underdoped, superconducting sample, with T$_c$ = 13 K, for $x=0.10$ (UD2).
    \item Maximally doped, superconducting sample, with T$_c$ = 27 K, for $x=0.18$ (OP).
\end{enumerate}
In this experiment, we explored a wider temperature range, from room temperature down to 2 K. We also used several photon wavelengths, $\lambda=$ 488 nm (blue), 532 nm (green), and 660 nm (red) to study resonance effects, but only linear polarizations (see Fig.~\ref{pol_config45} in Appendix \ref{selrul}. The results are compatible with those obtained in the preliminary test (see Fig.~\ref{Green_4config} in Appendix~\ref{Raman_supplemetary}).

Figure~\ref{RamanC_Tdep}(a) shows the temperature dependence in the antiferromagnetic compound \CCOC{} in the low- (up to the phonon Debye energy 85 meV) and intermediate (85-200 meV) energy-shift region from 21 K up to room temperature. Measurements correspond to the A$_{1g}$+B$_{1g}$ symmetry from linear polarized light selection rules, and were performed using the ``Grenoble setup”, with a red laser to achieve improved energy resolution.
On the other hand, in Fig.~\ref{RamanC_Tdep}(b), it is shown, using the ``Paris-setup", the Raman scattering from circularly polarized light in antiferromagnetic \CCOC, as a function of temperature from 30 K up to room temperature and with crossed polarization, having B$_{1g}$+B$_{2g}$ character.
The feature at around 400 meV stays at roughly the same position and almost the same intensity from 30K to 100 K, but starts to soften and lose intensity at 200 K, approaching T$_N$ ($\sim$260 K), and furthermore at 300 K. The overall softening is of about 20 meV from 397(9) meV at 30 K to 377(9) meV at 300 K.

\begin{figure*}[!htb]
	\includegraphics[width=1\linewidth]{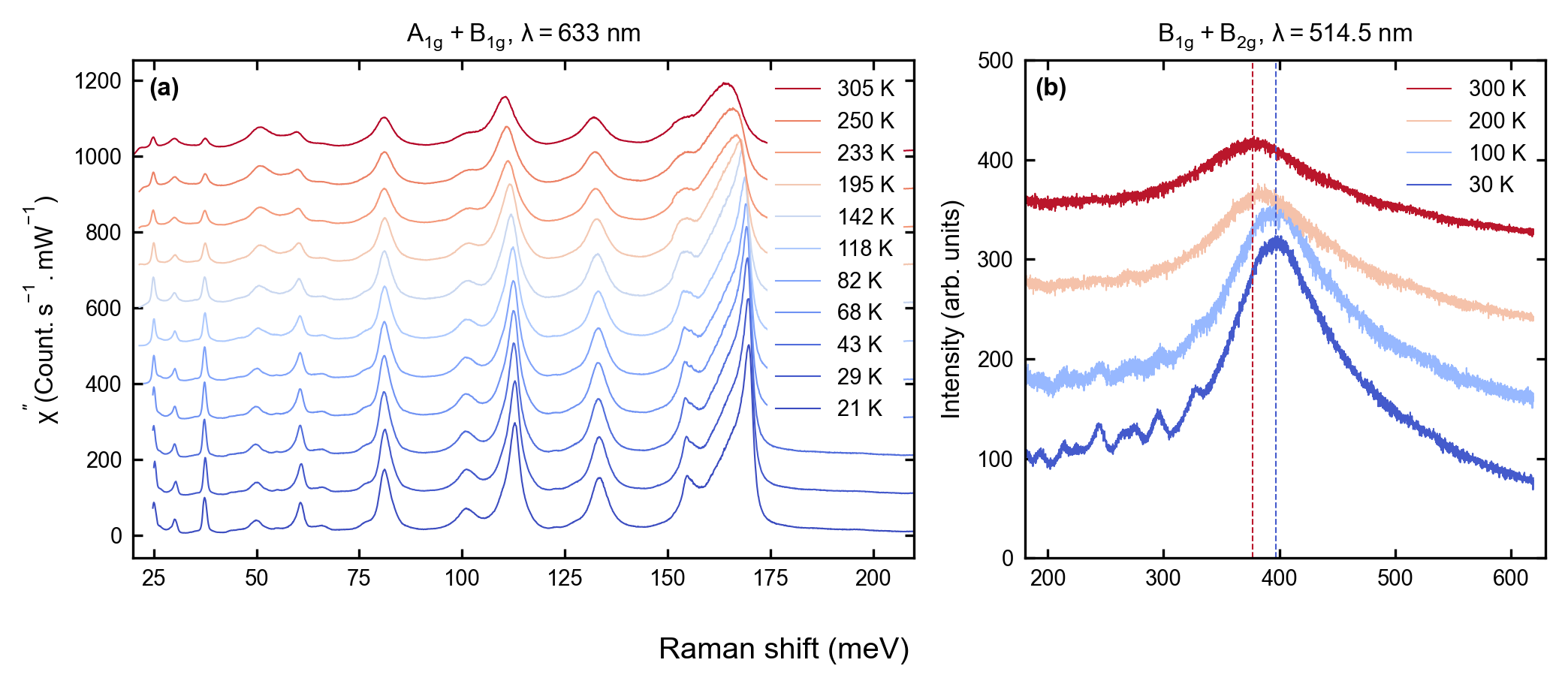}
	  \caption{\label{RamanC_Tdep} Temperature dependence of the Raman spectra in the 
      undoped \CCOC. (a) Raman scattering of the A$_{1g}$+B$_{1g}$ symmetry from linearly polarized light, from 21 K up to room temperature, using a red laser for improved energy resolution.   
      (b) Raman scattering of the B$_{1g}$+B$_{2g}$ symmetry from circularly polarized light, from 30 K up to room temperature using a green laser.
      \textit{See Appendix \ref{Raman_supplemetary} for the corresponding measurements with crossed linear polarisation with A$_{2g}$+B$_{2g}$ character, and uncrossed polarisation with A$_{1g}$+A$_{2g}$ character at the same temperatures. Red and blue lines mark the bimagnon energy at room temperature and 30~K, respectively.}}
\end{figure*}

\begin{figure*}[!htb]
   \centering
   \includegraphics[width=1 \textwidth]{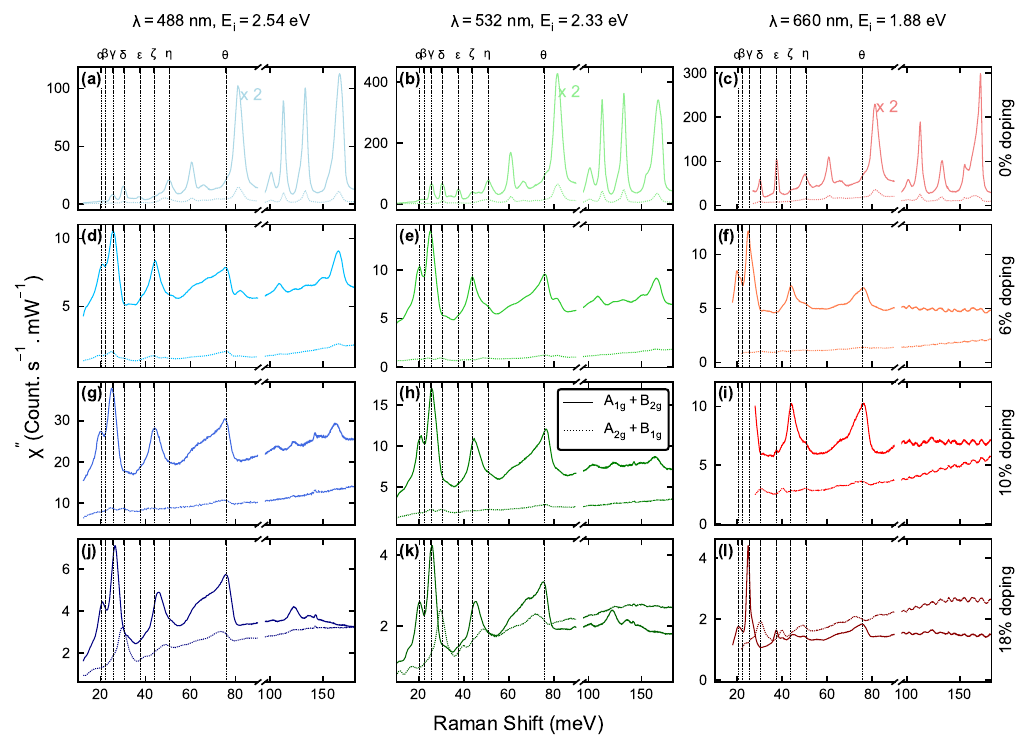}
   \caption{Phonon symmetry analysis at 2 K for all measured doping levels and laser wavelengths. The first, second, and third columns correspond to the blue, green, and red lasers, respectively.  The first, second, third, and fourth rows correspond to Na-doping levels of $x = 0,\, 0.6,\, 0.10,\, 0.18$, respectively.
   The diagonal marks on the x-axis indicate an axis break located around $\omega_D$, marking the upper limit of the one-phonon spectrum. The solid lines correspond to the $A_{1g}+B_{2g}$ symmetry while the dotted lines corresponds to the $A_{2g}+B_{1g}$.
   In panels (a),(b), and (c), the spectra to the left of the axis break were multiplied by 2 for both symmetries.
   Dashed lines denote the Raman-active modes ($A_{1g}$ labeled as $\gamma$ and $\epsilon$) and the one-phonon Raman-forbidden modes, as calculated for the undoped compound.}
   \label{Raman_doped}
\end{figure*}

The results at the lowest temperature (2 K) in the low- (up to the phonon Debye energy 85 meV) and intermediate- (85-200 meV) energy-shift region, with all photon wavelengths, are shown in Fig.~\ref{Raman_doped}.
The results are again consistent with those observed in the preliminary experiment on the undoped compound. However, the modes observed in the intermediate-energy region decrease strongly with both doping and incident photon energy, particularly for red photons ($\lambda=$ 660 nm), while the modes corresponding to expected Raman-active phonons remain visible and increase in relative intensity.  

\begin{figure*}[t]
    \centering
        \includegraphics[width=1\textwidth]{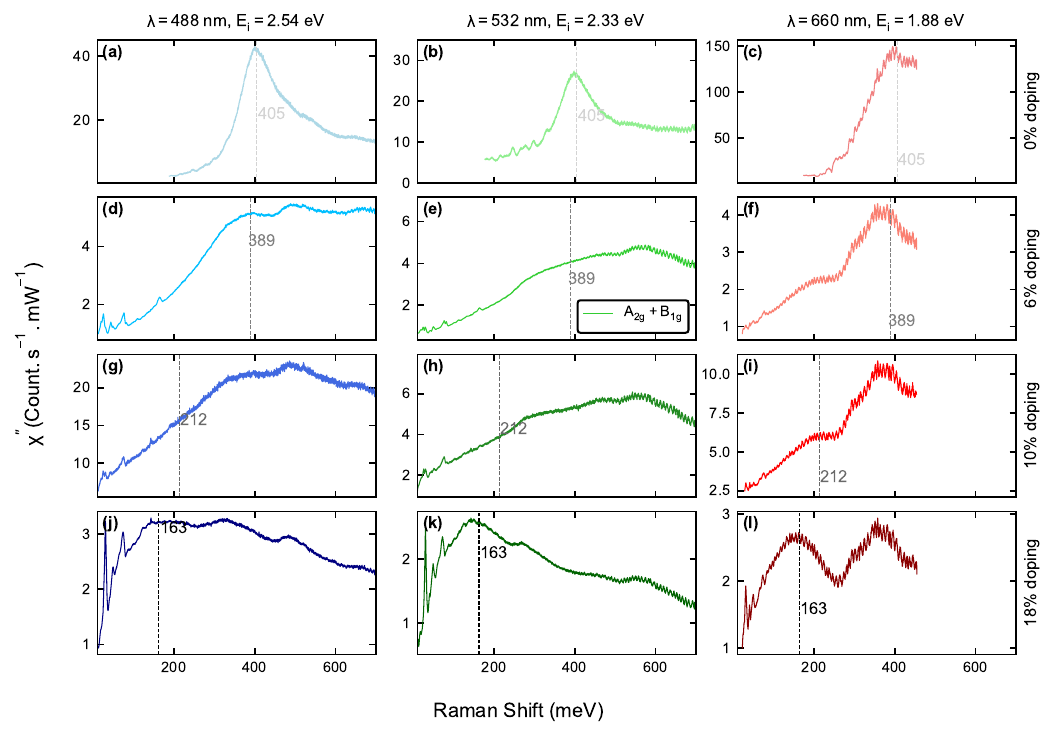}
        \label{B1g_bimagnon}
    \caption{Laser and doping dependence of the bimagnon at 2 K shown in the $A_{2g} + B_{1g}$ channel. The first, second, and third columns correspond to the blue, green, and red lasers, respectively.  The first, second, third, and fourth rows correspond to Na-doping levels of $x = 0,\, 0.6,\, 0.10,\, 0.18$, respectively. Vertical dashed lines give the position of the bimagnon as determined from the $B_{1g} - B_{2g}$ signal shown in Fig.~\ref{B1gmB2g} with the protocol detailed in Appendix \ref{Bimagnon_energy_det}. The position values are also reported in Table \ref{bi-mag_energies}.}
    \label{bimagnon_dop_dep}
\end{figure*}

Figure~\ref{bimagnon_dop_dep} shows the dependence of the high-energy region $A_{2g} + B_{1g}$ signal with doping and incident photon wavelength and polarisation, using the ``Grenoble-setup" at 2 K.
Since no $A_{2g}$ modes are expected, and because in the preliminary experiment, using the ``Paris-setup" with the full polarization analysis including the circular one, we confirmed the pure $B_{1g}$ character in the undoped compound, we can hypothesize that the signal has a pure $B_{1g}$ character. 

We note that using the red laser (1.88 eV), particularly in Fig.~\ref{bimagnon_dop_dep}(c), in the high-energy-shift region, above $\sim 200$~meV, a very broad and intense fluorescence signal arises, also centered around $\sim 400$~meV, which hides the inelastic one (see Fig.~\ref{fluo} in Appendix \ref{Raman_supplemetary}). 

\section{Discussion}
\subsection{Phonons}

\begin{figure*}[!htb]
   \centering
   \includegraphics[width=1 \textwidth]{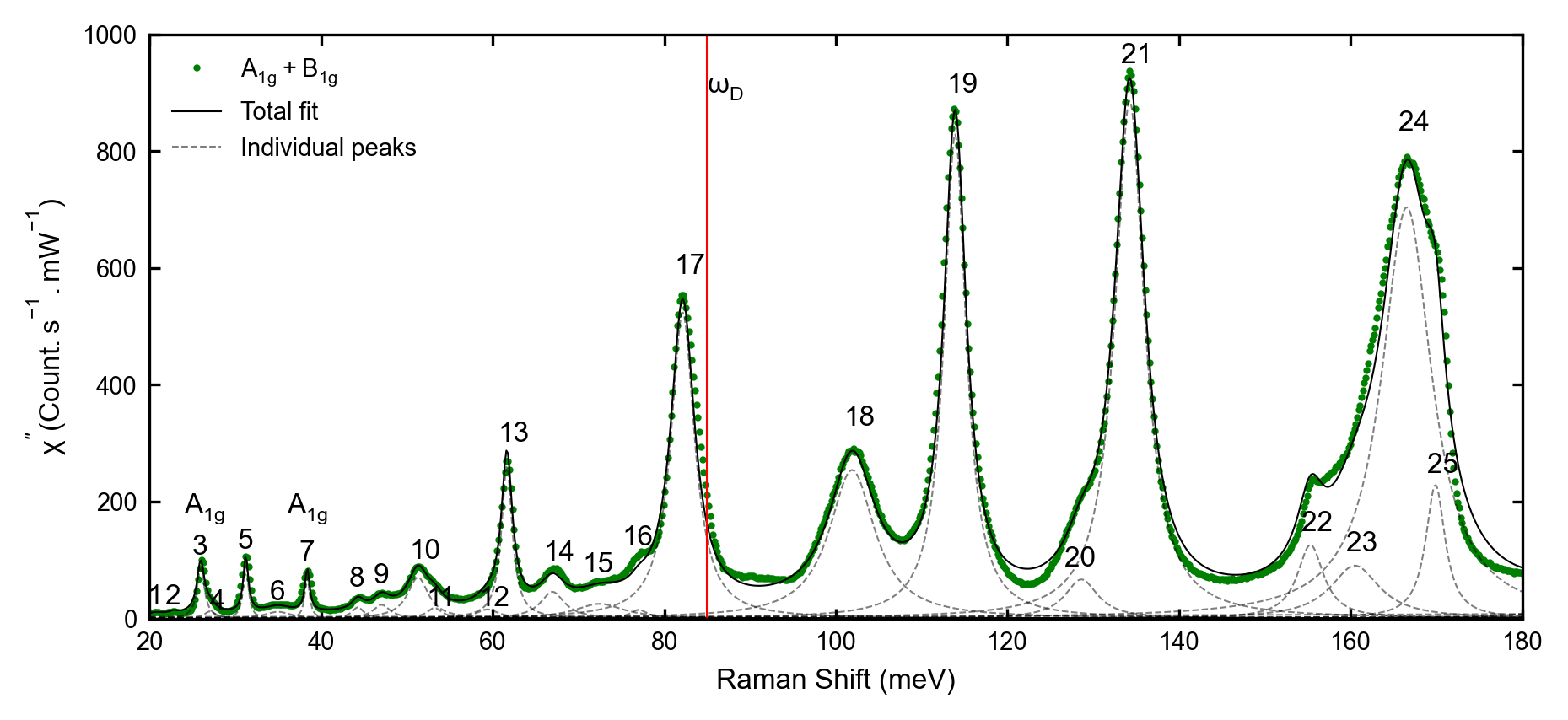}
   \caption{Raman spectra using the green laser ($\lambda = 532$ meV, $E_i = 2.33$ eV) of the $A_{1g}+B_{1g}$ symmetry of the undoped \CCOC\ at $T=21$ K. The solid line represents the total fit, while the dashed line represents the individual peaks fitted by the Lorentzian model. The red vertical line marks the Debye energy ($\omega_D$), representing the limit of the single-phonon spectrum.}
   \label{undoped_forest}
\end{figure*}

To assign the origin of the excitation that gives the main inelastic scattering signal in the energy region of 20 meV to 200 meV, we fitted the observed peaks using Lorentzian functions, in particular for the Raman signal collected using the green laser, where we have very good resolution on a freshly cleaved surface. The fit is presented in Fig.~\ref{undoped_forest} for the undoped \CCOC.

\begin{table*}[!htb]
    \centering
    \begin{tabular}{c|c|c|l || c|c|c|l}
    \hline \hline
        peak & energy  & FWHM  & mode
            & peak & energy &  FWHM & mode \\ 
        nb & (meV) & (meV) & 
            & nb &  (meV) &  (meV) & \\ \hline
            
            \hline
        1 & 20.72(7) & 1.47 & $A_{2u}(1)$ 
            & 14 & 66.97(17) & 3.43 & $A_{1g}(1) + A_{1g}(2)$ \\ [0.2cm]
           
        2 & 22.79(5) &  1.00 & $A_{2u}(2)$ 
            & 15 & 72.47(35) & 6.92 & $A_{2u}(1) + A_{2u}(3)$ \\ [0.2cm]
            
        3 & 25.97(5) & 1.05 & $A_{1g}(1)$ 
            & 16 & 77.03(12) & 2.44 & $E_u(4)$ or $A_{1g}(2) + A_{1g}(2)$ \\ [0.2cm]
            
        4 & 27.05(8) & 1.57 &$E_{u}(2)$ 
            & 17 & 82.15(17) & 3.36 & $E_u(4)$ at $(\pi,\pi)$ (from \cite{Lebert_2020})\\ [0.2cm]
            
        5 & 31.22(4) & 0.70 & $B_{u}$ 
            & 18 & 101.89(32) & 6.48 & $A_{2u}(3) + A_{2u}(3)$ \\ [0.2cm]
            
        6 & 34.96(18) & 3.66 & $E_{u}(1) + A_{2u}(1)$? 
            &  \multirow{2}{*}{19} & \multirow{2}{*}{113.89(15)} & \multirow{2}{*}{3.13} & $11 + 12 \rightarrow A_{1g}(1) + A_{1g}(1)+ E_{u}(1) + E_{u}(3) $  \\ [0.2cm]
            
        7 & 38.38(3) & 0.65 & $A_{1g}(2)$ 
            &  &  & & or $E_u(4) + E_u(4) - A_{2u}(2) $  \\ [0.2cm]
            
        8 & 44.28(9) & 1.70 & $E_{u}(3)$
            & 20 & 128.61(24) & 4.70 & $E_u(4) + A_{2u}(3)$\\ [0.2cm]
            
        9 & 47.05(12) & 2.45 & $E_{u}(2) + A_{2u}(1)$
            & 21 & 134.25(21) & 4.2 & $2*14 \rightarrow 2*(A_{1g}(1) + A_{1g}(2)) $ \\ [0.2cm]
            
        10 & 51.29(14) & 2.96 & $A_{2u}(3)$ 
            & 22 & 155.31(19) & 3.73 & $E_u(4) + E_u(4)$ \\ [0.2cm]
            
        11 & 53.22(11) & 2.20 & $A_{1g}(1) + A_{1g}(1)$? 
            & 23 & 160.5(35) & 6.98 & ?? \\ [0.2cm]
            
        12 & 59.4(16) & 3.23 & $E_{u}(1) + E_{u}(3)$ 
            & 24 & 166.53(35) & 7.10 & $E_u(4) + E_u(4)$ at $(\pi,\pi)$ \\ [0.2cm]
            
        13 & 61.65(7) & 1.53 & $B_{u} + B_{u}$ 
            & 25 & 169.88(14) & 2.76 & ??  \\ [0.2cm]

        \hline \hline
        
    \end{tabular}
    \caption{Peak energies from the fit of Fig.~\ref{undoped_forest} and assigned modes.
    (?) means there is large uncertainty about the assigned character due to other possible two-phonon combinations; see the text for details. (??) means an undetermined mode.}
    \label{forest_peak}
\end{table*}

The peak positions from the fit of Fig.~\ref{undoped_forest} are summarized in Table \ref{forest_peak}, where we indicate, when possible, their character from the comparison of the same peak in different polarizations (see, \textit{e.g.} Fig.~\ref{Raman_doped} for $A_{1g} + B_{2g}$ and $A_{2g} + B_{1g}$). 

The peak positions below $\sim 85 $ meV are tabulated with expected single-phonon modes from theory and previous infrared spectroscopy experiments, as shown in Tables~\ref{phonon_summary} and \ref{phonon_summary2}. 

We note that we observe the expected Raman-active $A_{1g}$ modes, but also all the other single-phonon modes, with the notable exception of the Raman-active $E_{g}$ modes. The latter are, in fact, expected in another geometry for the light scattering (with $\mathbf{k_i}$ perpendicular to the crystal axis $c$) \footnote{In any case, according to our calculations, even if present, the $E_{g}$ phonon modes would be expected at energies either below our low energy cutoff ($E_{g}(1)$) or possibly suprerposed to the $A_{1g}(1)$ ($E_{g}(2)$).}. 

\begin{table*}[!htb]
    \centering
    \begin{tabular}{|m{0.08\textwidth}
                    |m{0.1\textwidth}
                    |m{0.1\textwidth}
                    |m{0.1\textwidth}
                    |m{0.1\textwidth}
                    |m{0.1\textwidth}
                    |m{0.3\textwidth}
                    | }
        \hline \hline
       \multirow{2}{*}{Character}& \multirow{2}{*}{Description} &  \multicolumn{4}{c|}{Energy (meV)}  &  Displacement pattern \\ \cline{3-6}
      &   &  Hybrid DFT \cite{Patterson2008} & Shell model \cite{d2013phonon}  & DFT (this work)   & Raman fit (this work) & \\ \hline\hline
 \multirow{2}{*}{$A_{1g}$}      & Cu-Cl and Cu-Ca out-of-phase stretching     & 22.81 & 25.4 & 24.62 & 25.97(5) &\parbox[c]{1\linewidth}{\begin{center}\includegraphics[width=1\linewidth]{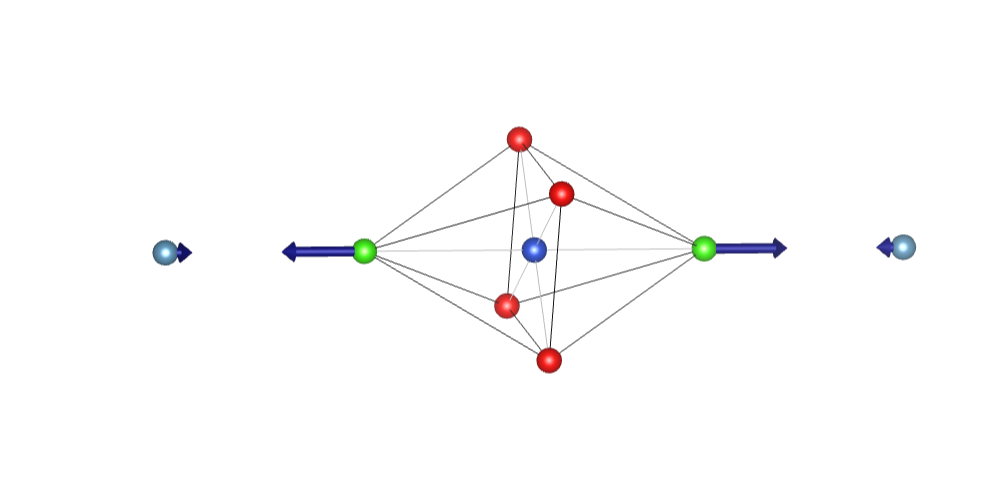}\end{center}}\\ \cline{2-7}

&        Cu-Cl and Cu-Ca  in-phase stretching   & 40.42 & 30.7 & 37.6 & 38.38(3) &\parbox[c]{1\linewidth}{\begin{center}\includegraphics[width=1\linewidth]{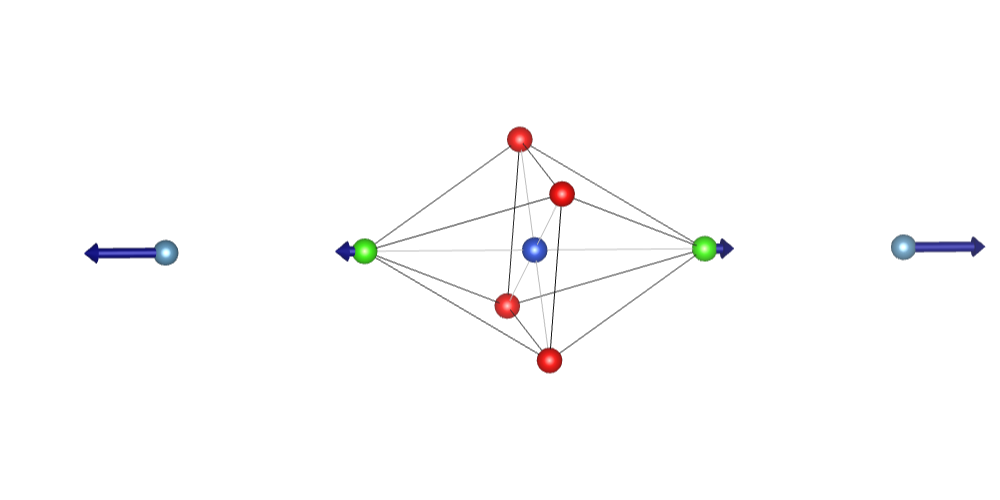}\end{center}}\\  \hline\hline
 \multirow{2}{*}{$Eg$}      &    Cu-Cl bend + Ca motion        & 16.49  &  &16.71 &  &\parbox[c]{1\linewidth}{\begin{center}\includegraphics[width=1\linewidth]{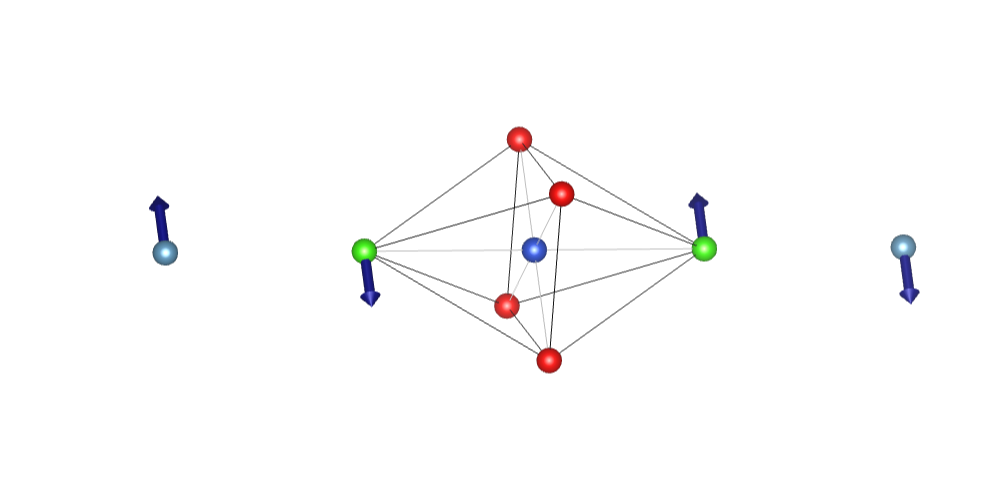}\end{center}}\\ \cline{2-7}
  &   Cu-Cl bend                         & 23.19 & - & 24.77 & & \parbox[c]{1\linewidth}{\begin{center}\includegraphics[width=1\linewidth]{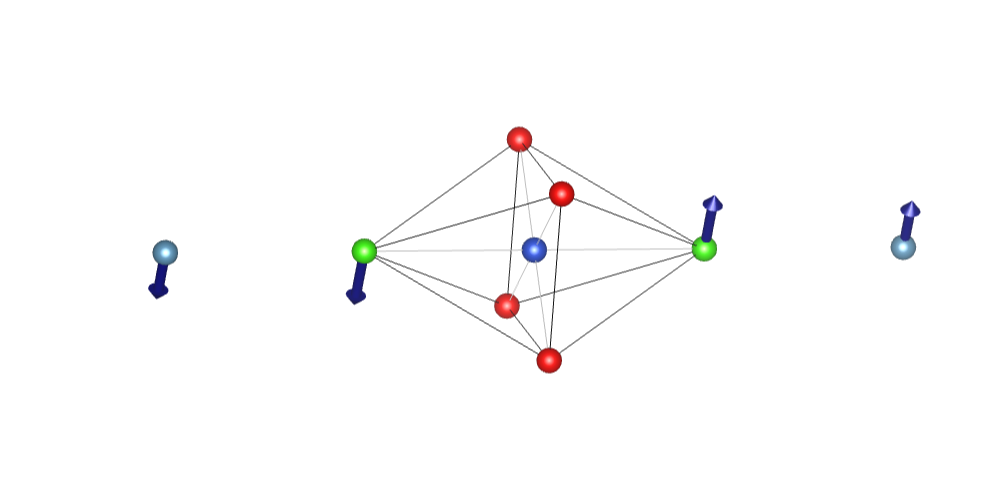}\end{center}}\\   \hline\hline
$B_u$   &Cu-O out-of-phase buckling       & 29.14 & - & 29.9 &  31.22&\parbox[c]{1\linewidth}{\begin{center}\includegraphics[width=1\linewidth]{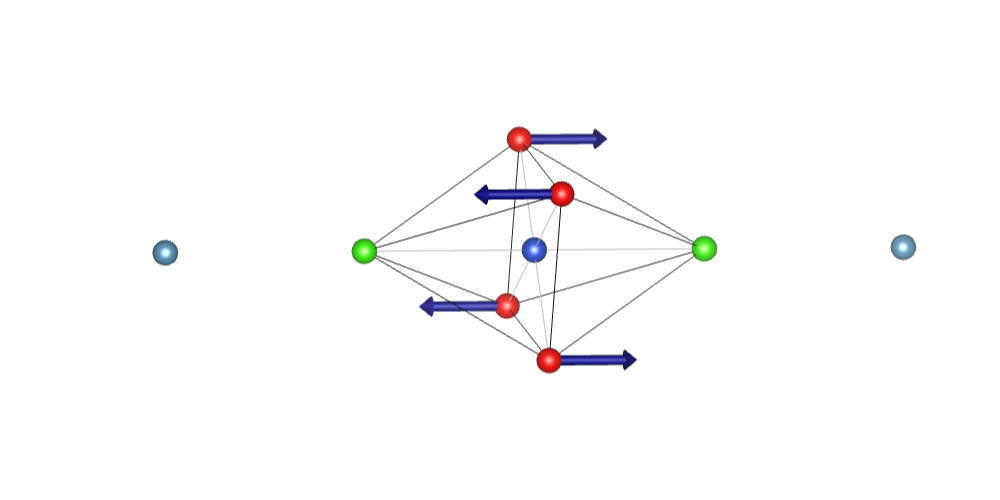}\end{center}} \\  \hline \hline
 \end{tabular}
    \caption{Expected energies for the Raman optical modes in undoped \CCOC, from theoretical calculations, including an optically silent mode ($B_{u}$). Owing to the (D$_{4h}$) symmetry of \CCOC, only the $A_{1g}$ should be Raman active in our geometry, while $B_u$, which is supposed to be optically silent, is activated in our measurements (see text). The mode displacement patterns, in this and the following Table \ref{phonon_summary2}, correspond to the ones calculated by DFT in this work. The sketch is rotated by 90$^\circ$ with respect to the standard orientation of Fig.~\ref{ccoc structure}, turning the $c$ crystal axis horizontal, to optimize the table layout. Each ion species is represented with the same color code as in Fig.~\ref{ccoc structure}, but with spheres of fixed radius to better compare the displacement.}
    \label{phonon_summary}
\end{table*}

\begin{table*}[!htb]
    \centering
    \begin{tabular}{|m{0.08\textwidth}
                    |m{0.1\textwidth}
                    |m{0.1\textwidth}
                    |m{0.1\textwidth}
                    |m{0.1\textwidth}
                    |m{0.06\textwidth}
                    |m{0.1\textwidth}
                    |m{0.28\textwidth}
                    |}
        \hline \hline
       \multirow{2}{*}{Character}& \multirow{2}{*}{Description} &  \multicolumn{5}{c|}{Energy (meV)}  &  Displacement pattern\\ \cline{3-7}
      &   &  Hybrid DFT \cite{Patterson2008} & Shell model \cite{d2013phonon}  & DFT (this work)  & IR \cite{zenitani2005IR}  & Raman fit (this work) & \\ \hline\hline

\multirow{3}{*}{$A_{2u}$}      &   Cu-Cl stretching + Ca motion   & 19.47 &  - & 18.74 &  - &  20.72(7)&  \parbox[c]{1\linewidth}{\begin{center}\includegraphics[width=1\linewidth]{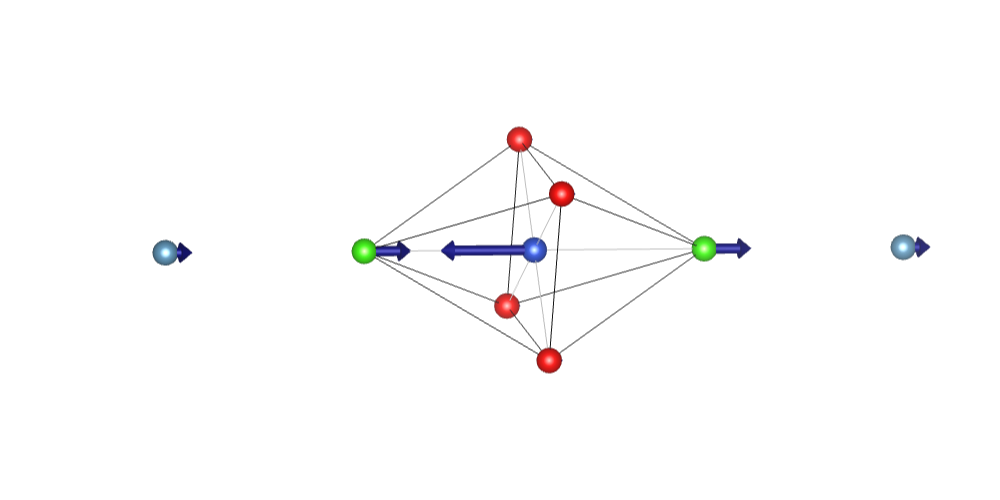}\end{center}} \\  \cline{2-8}
  &      Cu-O in-phase buckling         & 22.19 & - & 21.52 & - & 22.79(5) &\parbox[c]{1\linewidth}{\begin{center}\includegraphics[width=1\linewidth]{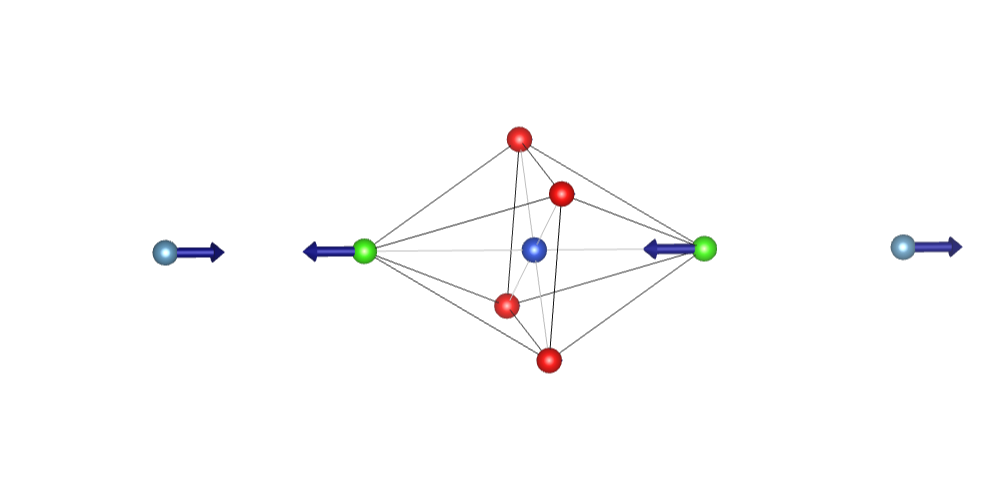}\end{center}}\\  \cline{2-8}
  &  Cu-O in-phase buckling         & 50.46 & - & 45.54 & 53.35*& 51.59(14) &\parbox[c]{1\linewidth}{\begin{center}\includegraphics[width=1\linewidth]{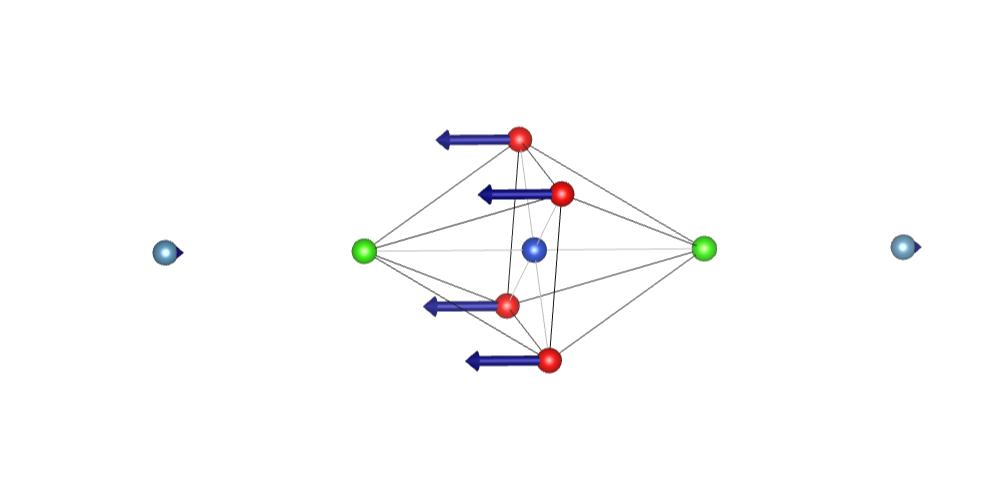}\end{center}}\\ \hline\hline
\multirow{4}{*}{$E_u$}        &   Cu-Ca bending                         & 14.26 & 12.7 & 15.49 & - & -&\parbox[c]{1\linewidth}{\begin{center}\includegraphics[width=1\linewidth]{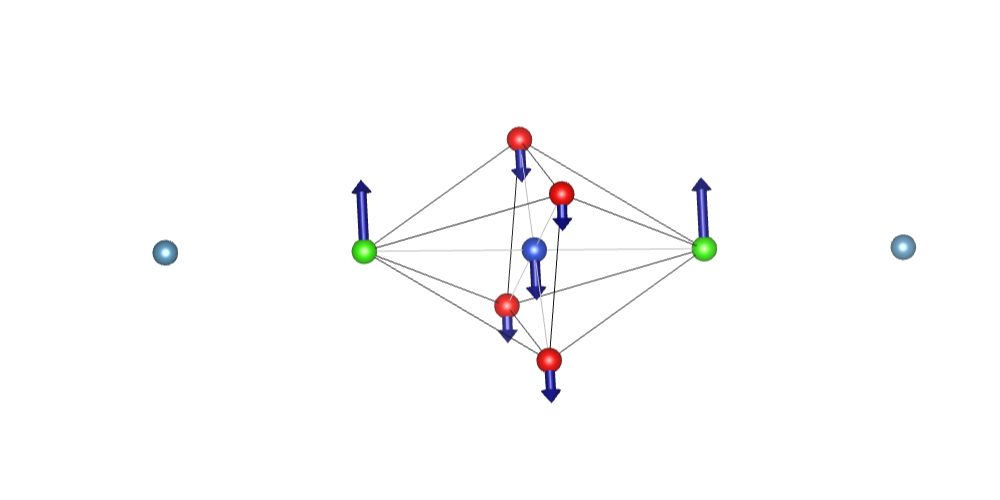}\end{center}}\\   \cline{2-8}
  &      CuO$_4$Cl$_2$   + Ca motion                & 26.04  & 24.53 & 25.32 &-&27.05(8) &\parbox[c]{1\linewidth}{\begin{center}\includegraphics[width=1\linewidth]{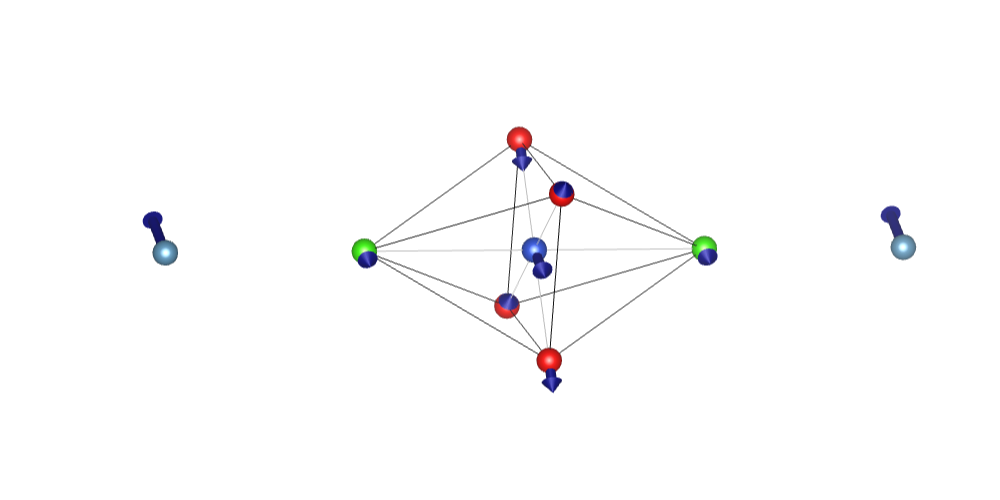}\end{center}}\\ \cline{2-8}
  
   &    Cu-O in-plane bend                 & 41.41 & 43.5 & 44.45 & 43.42&44.28(9) &\parbox[c]{1\linewidth}{\begin{center}\includegraphics[width=1\linewidth]{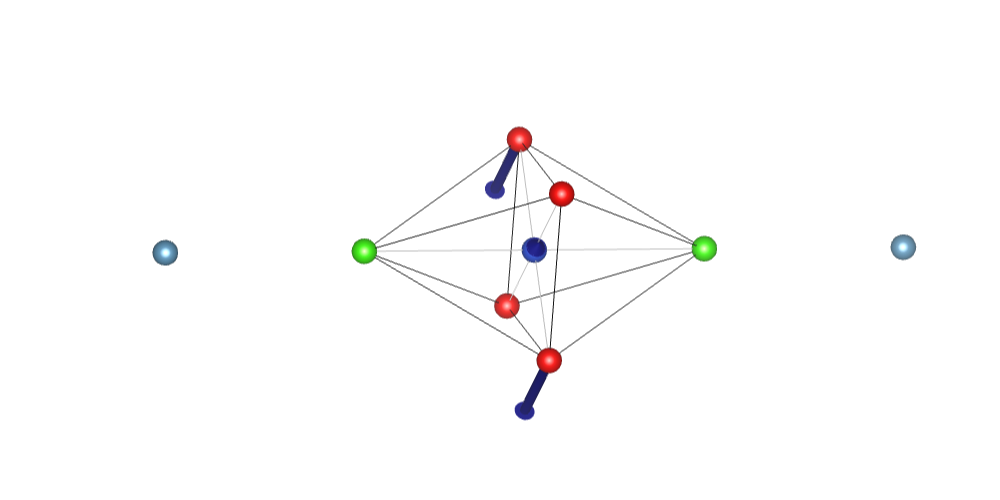}\end{center}}\\  \cline{2-8}
    &    Cu-O stretching                    & 76.26 & 74.5 & 74.18& 75.93&77.03(12)&\parbox[c]{1\linewidth}{\begin{center}\includegraphics[width=1\linewidth]{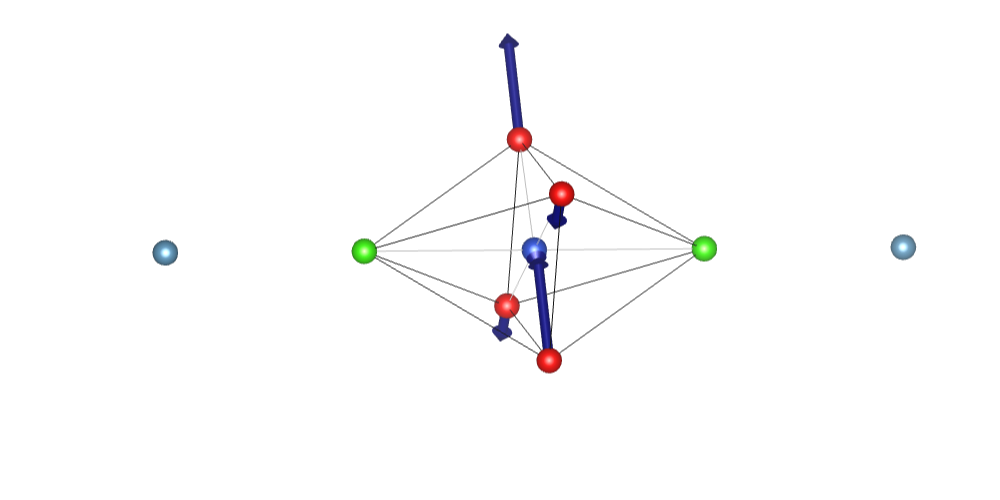}\end{center}} \\ \hline \hline

 \end{tabular}
    \caption{Expected energies for the infrared optical modes in undoped \CCOC, from theoretical calculations and infrared spectroscopy.\\
    *While this mode is not confirmed in \cite{zenitani2005IR} as an $A_{2u}$, it has a similar energy to the one calculated in Hybrid DFT \cite{Patterson2008}.}
    \label{phonon_summary2}
\end{table*}

We note that, while the observed $A_{1g}$ modes show a sharp and symmetric profile, the other observed modes, which are not supposed to be Raman-active but have energies corresponding to expected single-phonon modes, show broader and more asymmetric profiles.

Most of the peaks that do not correspond to single modes can match the expected energy and symmetry of two-phonon modes.
Indeed, we calculate all the possible combinations for two-phonon processes that are Raman active (see Table \ref{two_phonon_Raman} in Appendix \ref{Raman_supplemetary}). The calculations are based on the measured phonon energies, except for E$_u$(1) and E$_g$(1), which are below the measured range.
Also, for these modes, the energy positions of the peaks, as labeled in Fig.~\ref{undoped_forest}, are listed with their character in Table \ref{forest_peak}.
There is a large number of possible combinations for multiple phonon scattering. Therefore, in order to assign a character to these peaks, we choose the nearest corresponding combination, as shown in Tables \ref{Raman_active_2} and \ref{two_phonon_Raman} in Appendix~\ref{Appendix:twoph}. When different combinations match a certain peak energy, we prioritized choosing the combination of two intense phonons or two Raman-active phonons.
Except for two peaks, all the non-single phonons match well with the two-phonon process and, in some cases, it might correspond to even higher-order harmonics, as shown in Table \ref{forest_peak}. This just provides an empirical approximation, whereas a full calculation of all allowed combinations would yield the partial density of states for the multi-phonon processes, which indeed seems to be the case, given their broad and asymmetrical lineshapes. However, the double-phonon mode seems to be the dominant process, hence a good approximation of the effect. 

\begin{table}[!htb]
    \centering
    \begin{tabular}{c|c|l||c|c}
    \hline \hline
        \multicolumn{3}{c||}{Undoped, x=0} & \multicolumn{2}{c}{Maximally doped} \\ \hline
        peak & energy (meV) & mode          & peak & energy (meV)\\ \hline
        1 & 20.12(7) & $A_{2u}(1)$             & $\alpha$ & $\sim$21 \\ [0.2cm]
        2 & 22.17(5) & $A_{2u}(2)$             & $\beta$  & $\sim$22 \\ [0.2cm]
        3 & 25.43(5) & $A_{1g}(1)$             & $\gamma$ & $\sim$26.5 \\ [0.2cm]
        4 & 26.54(8) & $E_{u}(2)$              & - & shoulder? \\[0.2cm]
        5 & 30.33(4) & $B_{u}$                 & $\delta$    & $\sim$29.5 \\  [0.2cm]
        7 & 37.50(3) & $A_{1g}(2)$             & $\epsilon$ & weak \\ [0.2cm]
        8 & 43.63(9) & $E_{u}(3)$              & $\zeta$ & $\sim$45.5 \\[0.2cm]
        10 & 50.68(8) & $A_{2u}(3)$            & $\eta$ & $\sim$50 \\ [0.2cm]
        16 & 75.71(5) & $E_u(4)$ & $\theta$ & $\sim$75.5\\ \hline \hline
    \end{tabular}
    \caption{Comparison between phonons in undoped \CCOC\ and doped \NaCCOC. Values for the undoped sample were obtained by fitting the data reported in Fig.~\ref{Raman_doped}. For the maximally doped sample, values were approximated from the energy position of the peak maximum, as reliable fitting is hindered by substantial spectral broadening and a large electronic background.}
    \label{forest_peak_doped}
\end{table}

Possible origins of forbidden single-phonon mode activation,  in particular \IR\ ones, as well as multi-phonon modes and their harmonics, include impurity effects and additional crystal symmetry breaking in the compound. 

Another possibility is resonance effects. 
Indeed, the activation of Raman forbidden modes in cuprates has been reported and discussed as a resonance of the incident energy with the charge transfer energy corresponding to the O(p)-Cu(d) bond, in a wide range of cuprates families, such as SrCuO$_2$ in \cite{popovic2001}, YBa$_2$Cu$_3$O$_y$ (YBCO), La$_{2-x}$Sr$_x$CuO$_4$ (LSCO) \cite{Sugai1989}, Bi$_2$Sr$_2$Ca$_{1-x}$Cu$_2$O$_{8+\delta}$ (Bi2212) and Bi$_2$Sr$_{2-x}$La$_x$CuO$_{6+\delta}$ (Bi2202) in \cite{sugai2003}, Pb$_2$Sr$_2$PrCu$_3$O$_8$ \cite{Reedyk_1994}, as well as in electron-doped cuprates \cite{sugai2004}. 

The resonance induces Fr\"{o}hlich interaction (see later), leading to the appearance of \IR\ active modes, which are dipole-forbidden in Raman scattering. It also allows us to be sensitive to second-order modes, including overtones and two-phonon scattering. It can also enable the appearance of modes from different propagating vectors and is not restricted to the zone center \cite{sugai2003}.

In \CCOC, the charge transfer is measured in \cite{Lee2006}, to be equal to 2.2 eV, which is close to the green laser energy of 2.33 eV.
To verify the hypothesis of an activation of forbidden modes by resonance effects, we measure the Raman response of our samples with different photon energies, as shown in Fig.~\ref{Raman_doped}, where we report data with $h\nu=$2.54 eV (wavelength $\lambda=$ 488 nm, blue light, Fig.~\ref{Raman_doped}(a)), 2.33 eV (532 nm, green light, Fig.~\ref{Raman_doped}(b)) and 1.88 eV (660 nm, red light, Fig.~\ref{Raman_doped}(c)).
Lines from top down corresponds to doping of $x=0.0$ (top), $x=0.06$, $x=0.10$, and $x=0.18$ (bottom), and on each panel we compare $A_{1g}+B_{2g}$ to $A_{2g}+B_{1g}$, for the given photon energy and sample doping. 

Indeed, we clearly observe that the relative intensity of the Raman-active $A_{1g}(2)$ mode at $\sim$38 meV in undoped \CCOC{}, increases relative to the forbidden ones, as the excitation energy is tuned towards the red laser. This behavior strongly supports the interpretation in terms of resonance, and data in the $A_{1g}+B_{1g}$ channel as reported in Fig.~\ref{undoped_forest} show the same for the $A_{1g}(1)$ mode. 

This resonance effect can therefore explain the appearance of the non-Raman active modes and the appearance of the bond-stretching mode at the zone boundary (as peak 17 in Table \ref{forest_peak}, at 82.5 meV), as well as the scattering of higher-order modes.
Since the resonance is related to the Cu-O bond, we selectively enhance the atomic vibrations in the CuO$_2$ planes \cite{Reedyk_1994, sugai2003}. Therefore, we get similar or greater intensity of the related modes, $A_{2u}$, $E_u$, and $B_u$, compared to $A_{1g}$ modes corresponding to the Cu-Cl and Cu-Ca bonds, as presented in detail in Fig.~\ref{undoped_forest} and Table \ref{forest_peak}.  Moreover, the intensity loss of these modes when using the red laser with an $h\nu=1.88$ eV (Fig.~\ref{Raman_doped}(c,f,i,l)) is explained by exciting the resonance regime with an incident photon energy of 1.88 eV, smaller than the charge-transfer one, when using red light in the undoped compound. 

Concerning the doping effects, in the energy region above the Debye energy (85 meV), the two-phonons rapidly lose their strong intensity, seen in the undoped, adding as low as $x=0.06$ of holes (sample UD1). The two-phonon peaks then become weaker and weaker with doping, almost disappearing at maximum doping. Also the $E_u(4)$ at $(\pi, \pi)$ mode seems to lose its intensity in favor of $E_u(4)$ at the centre zone $\Gamma$.

The energy region below the Debye energy (85 meV) contains single- and two-phonon modes, which are generally well separated in the undoped compound. The broadening of these peaks with doping does not allow us to distinguish between them. However, one can still identify some peaks highlighted in Fig.~\ref{Raman_doped} by vertical dashed lines.
The corresponding energy positions are shown in Table \ref{forest_peak_doped} for both the maximally doped and the undoped sample. 

Peaks $\alpha$ and $\beta$ correspond to the $A_{2u}$ \IR\ active mode, which appears with $A_{1g}$ symmetry, similar to what has been observed in the undoped sample. Peaks $\gamma$ correspond to $A_{1g}(1)$ phonon.\\
Peak $\delta$ corresponds to the $B_u$ phonon, which appears in $A_{1g}$ symmetry in the undoped sample, but in the $x=0.06$ and $x=0.10$ doping, it appears to be absent or very weak. An excitation at a similar energy emerges at the maximal doping ($x=0.18$) in $B_{1g}$ symmetry.

Peak $\epsilon$ corresponds to the $A_{1g}(2)$, which is very weak in all the doped samples.
Peak $\zeta$ and $\theta$ corresponds to $E_u$ modes, \IR\ active, appearing in $A_{1g}$ symmetry, similar to the undoped case.
Peak $\eta$ is another $A_{2u}$ \IR\ active mode, present in $A_{1g}$ symmetry in the undoped. In the underdoped samples ($x=0.06 \, , 0.10$), it appears as a shoulder in the same symmetry ($A_{1g}$), then, in the maximally doped samples, it appears in the $B_{1g}$ symmetry, with similar behavior of peak $\delta$.

To explain the resonance effect, we have to take into account the Fr\"{o}hlich interaction, which refers to the long-range interaction between electrons and phonons and reflects a strong electron-phonon coupling \cite{Reedyk_1994, sugai2003}. These interactions are mediated by the microscopic electric field caused by the motion of vibrating ions. Introducing charge carriers screens this electrical field, causing these interactions to decrease \cite{sugai2003}. This is in agreement with the doping effect observed in \NaCCOC, where we see that two-phonon scattering is weakened upon doping. However, we still observe the first-order non-Raman active modes.

\subsection{Bimagnon}

As shown in Fig.~\ref{fig:CCOC_per_ramanpol}(a) and Fig.~\ref{RamanC_Tdep}(b), and detailed in the previous Sec.~\ref{Results}, the high-energy Raman shift part of our Raman spectra in antiferromagnetic \CCOC{} shows a signal around 400 meV, and that signal has a clear B$_{1g}$ character, according to the polarisation analysis reported in Fig.~\ref{fig:CCOC_per_ramanpol}(a).
The signal can be identified as Raman scattering from a bimagnon excitation, following previous reports in cuprates, where a similar signal in the same energy range was interpreted as coming from bimagnon excitations \cite{Lyons1988,PhysRevB.42.1045,Sugai1988}.
We also note, as reported above in Sec. \ref{Result2}, and shown in Fig.~\ref{RamanC_Tdep}, that this signal starts to soften and lose intensity at 200 K, and even more at 300 K, above T$_N\sim$260 K, which is compatible with an interpretation in terms of magnetic excitation. Details of the data analysis procedure we use to extract and model the B$_{1g}$ signal are given in Appendix \ref{Bimagnon_energy_det}.

The energy of the bimagnon in the antiferromagnetc phase, according to these results, it is of 397 $\pm$ 9 meV, which is 2.7(2)$\mathcal{J}$, using the Loudon-Fleury approach \cite{Fleury_Loudon_1968,Lyons1988,Devereaux_2007}, where $\mathcal{J}$ is the first neighbor super-exchange, and its value is reported in Ref.  [\onlinecite{PhysRevB.95.155110}] and [\onlinecite{Lebert_2023}], from a Heisenberg model fit of the magnon-dispersion.
The result is consistent with previous reports of a bimagnon energy in \LCO{}, also corresponding to approximately 2.7$\mathcal{J}$, using the same approach, which is valid when the excitation is less than 2 times the charge transfer gap $\Delta$. Indeed, in the present study, this is true, as a $\lambda$ = 514.5~nm photon corresponds to an energy of 2.41~eV, well below the value of 2$\Delta$ = 4.4~eV, with $\Delta$ being the gap energy.

It is worth noting that in Mott insulators, the bimagnon signal is enhanced when the excitation energy is of the order of the Mott gap energy \cite{Devereaux_2007}. In our case, this is replaced by the charge transfer gap $\Delta$ = 2.2~eV, with a behavior similar to that of the resonant phonons. Such behavior has also been seen in different cuprates \cite{Blumberg1994, Rubhausen1997, sugai2003, Tassini2008}. Indeed, Li \textit{et al.} \cite{Li_Letacon_2013} showed that the bimagnon has resonance properties depending on the doping, which coincides with an intraband transition rearrangement. The authors investigated the doping dependence of the bimagnon with doping in HgBa$_2$CuO$_{4+\delta}$ with Raman spectroscopy using two different excitation energies ($h\nu=$2.33 eV and $h\nu=$1.96 eV). They found that the bimagnon response was enhanced with a green laser ($h\nu=$2.33 eV) at lower doping levels. As the doping increased, the resonance shifted toward lower excitation energy (red laser with $h\nu=$1.96 eV), with a crossover occurring around 10$\%$ doping. Using ellipsometry, they identified two prominent intraband transitions: one centered around 1.2~eV, whose spectral weight increases with doping, and another around 2.9~eV, which was strongest in the low-doping regime.

We observe a similar effect on the bimagnon signal in doped \NaCCOC, as shown in Fig.~\ref{bimagnon_dop_dep}, and even more directly in Fig.~\ref{B1gmB2g}, where the B$_{1g}$ signal is isolated, as detailed in Appendix~\ref{Bimagnon_energy_det}.  The bimagnon energy positions, obtained from fitting that signal, are reported in Table~\ref{bi-mag_energies}. Figure~\ref{B1gmB2g} shows the change in the Raman scattering intensity in the B$_{1g}$ channel at the lowest temperature, where the bimagnon tends to lose intensity with doping and soften to lower energies. 
Hence, by using three different photon wavelengths, we can tune the energy to approximately optimize the resonance for each doping level.
In the undoped compound (antiferromagnetic), and for $n=0.06$ (strongly underdoped, not superconducting), the B$_{1g}$ signal is strongest using $h\nu= 2.54$ eV (blue laser), as shown in Fig.~\ref{bimagnon_dop_dep}(a) and  Fig.~\ref{bimagnon_dop_dep}(d), respectively (see also Fig.~\ref{B1gmB2g}, same panel, for a more clear determination of the B$_{1g}$ contribution as explained in Appendix \ref{Bimagnon_energy_det}.
Then, for both $n=0.1$ and $0.18$, the two dopings in the superconducting phase, the resonance becomes stronger for the red laser $h\nu= 1.88$ eV in Fig.~\ref{bimagnon_dop_dep}(i,l).
The signal loses intensity and broadens, while also showing a strong asymmetry and a large shift to low energy, as reported in Table~\ref{bi-mag_energies}, going from 405(6) meV in the antiferromagnetic phase to less than half of that value, at 163(8) meV for the maximum doping. 

We note that in our case, the situation is further complicated by a fluorescence signal that obscures the bimagnons at some incident photon energies (see Fig~\ref{fig:fluo}). For this reason, the optimal energy to observe the signal is a compromise between (i) the presence of the fluorescence signal, (ii) the bimagnon energy that softens and eventually exits the fluorescence region for the red laser at high doping, and (iii) the resonant incident photon energy for a given doping.


\begin{table}[!htb]
    \centering
    \begin{tabular}{c|c|c}
    \hline \hline
        Doping n & At photon $\lambda$ (nm) & energy (meV)  \\ \hline
        0.00 & 488 & 405(6)  \\ 
        0.06 & 488 & 389(40)  \\ 
        0.10 & 660 & 212(15) \\ 
        0.18 & 660 & 163(8) \\ 
        \hline \hline
    \end{tabular}
    \caption{Peak energies from a Lorentzian function fit of the main signal in Fig.~\ref{B1gmB2g}, at the photon energy that maximizes the resonance as indicated.}
    \label{bi-mag_energies}
\end{table}

\section{Conclusions}

In conclusion, we measured four different dopings, sampling the entire phase diagram of the oxychloride cuprate, from the antiferromagnetic parent compound to the maximally doped superconducting sample.

In addition to the expected Raman-active phonon modes, we detected several excitations that can be interpreted as single- and double-phonon modes, as well as their harmonics. 
Using different laser wavelengths, we can show that these forbidden modes are excited \textit{via} a resonant process due to strong electron-phonon coupling, as has already been reported for other cuprates \cite{Reedyk_1994,popovic2001,sugai2003,sugai2004}.

Finally, we observe a well-defined $B_{1g}$ mode in the antiferromagnetic sample, that can be interpreted as a bimagnon, and we follow its temperature and doping dependence, giving valuable information on multimagnon excitations in these cuprates and helping us better understand their contribution to the magnetic dynamics as seen, \textit{e.g.}, by resonant x-ray inelastic scattering (RIXS) \cite{Lebert_2023}.

\section*{Acknowledgment}
This work was supported by the European Research Council (ERC) under the European Union’s Horizon 2020 research and innovation program (Grant Agreement n◦ 865826).
For help on single-crystal diffraction, we thank Beno\^it Baptiste for the use of the X-Ray diffraction platform at IMPMC, as well as Olivier Leynaud and the X’Press engineering pool at Inst. N\'eel.  
We thank Didier Dufeu and the PMag platform for help in the magnetometry measurements, as well as Nathan Bujault, Felix Morineau, Carley Paulsen, and André Sulpice for additional magnetization measurements. We acknowledge help for sample synthesis at Institut NEEL from Stefan Schulte, Anne Missiaen, Murielle Legendre, Céline Goujon, and Pierre Toulemonde.
 
\bibliographystyle{apsrev}
\bibliography{references}

\newpage

\appendix

\begin{table*}[!htb]
    \centering
    \begin{tabular}{c|c|c|c|c} 
    \hline \hline
    Configuration & Active symmetry & Leakage    & Bimagnon \\ \hline
    $\overline{c}(a,a)c$    & $A_{1g}+B_{1g}$   & $A_{2g}+B_{2g}$               & present \\
    $\overline{c}(a,b)c$    & $A_{2g}+B_{2g}$   & $E_g(2)+A_{1g}+B_{1g}$        & absent \\
    $\overline{c}(ab,ab)c$  & $A_{1g}+B_{2g}$   & $A_{2g}+B_{1g}$               & absent \\
    $\overline{c}(ab,a\overline{b})c$    & $A_{2g}+B_{1g}$   & $E_g(1)+E_g(2)+A_{1g}+B_{1g}$ & present \\
    $\overline{c}(L,L)c$    & $A_{1g}+A_{2g}$   &  & absent \\
    $\overline{c}(L,R)c$    & $B_{1g}+B_{2g}$   &  & present \\ \hline \hline
    \end{tabular}
    \caption{Selection rules for D$_{4h}$ point group. $ab$ and $a\overline{\text{b}
    }$ are the  $(110)$ and $(1\overline{1}0)$ crystalline axes, from \cite{Buhot}.}
    \label{D4h_selection_rules}
\end{table*}

\section{Selection rules}\label{selrul}

The \CCOC\ has an $I4/mmm$ structure, and $D_{4h}$ point group. The atoms' Wyckoff positions are as follows: Cu($2b$), Cl($4e$), Ca($4e$), O($4c$). Only the $A_{1g}$ and $E_{g}$ symmetries corresponding to $4e$ sites are Raman active according to Bilbao Crystallographic Server calculations \cite{Bilbao_server}. For the bimagnon, it is known in cuprates that it appears in $B_{1g}$ symmetry \cite{Fleury_Loudon_1968, Devereaux_2007}. We also note that the $A_{2u}$ and $E_u$ modes are \IR\ active, and the $B_u$ mode is Hyper-Raman active, which is not feasible on our experimental setup.

\begin{figure}[h]
   \centering
   \includegraphics[width=0.45 \textwidth]{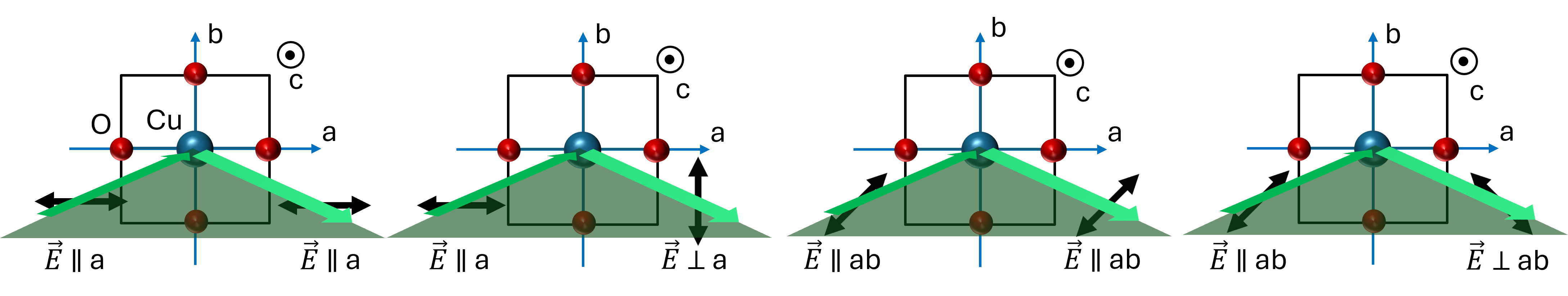}
   \caption{Incident-scattered polarization configurations with a $30^{\circ}$ scattering angle with respect to the $c$-axis: left $\overline{c}(a,a)c$; left center $\overline{c}(a,b)c$; right-center $\overline{c}(ab,ab)c$; right $\overline{c}(ab,a\overline{b})c$.
   Back-scattering configurations have both incident and scattered light along the $c$-axis and the same polarization scheme.}
   \label{pol_config45}
\end{figure}

The incident beam has a pointing vector making an angle of $0^\circ$ (preliminary experiment in backscattering geometry, Paris-setup) and $30^\circ$ (main experiment, Grenoble-setup) with the sample $c$-axis. 
However due to the strong refraction index inside the sample, the pointing vector can be considered as along \textit{c}-axis, even with the latter geometry. Fig.~\ref{pol_config45} shows the four incident-scattered polarization configurations aligned with the crystal axes along with the typical unit cell of the CuO$_2$ planes. The vibration direction of the electrical field is either parallel or perpendicular to the $a$-axis or to the $ab$ (110) axis. In this case, The Selection rules are summarized in Table~\ref{D4h_selection_rules} for the scattering geometry used in this work.

Consequently, based on Table \ref{D4h_selection_rules}, we should see two $A_{1g}$ modes corresponding to Ca and Cl vibrations, along with the bimagnon in $B_{1g}$ symmetry, in addition to possible leakage of $E_g$ modes.\\
The other possible Raman-active modes are second-order Raman scattering, corresponding to the two-phonon process, summarized in Table \ref{Raman_active_2}.

\section{Two-phonon Raman active scattering}\label{Appendix:twoph}

In Table \ref{Raman_active_2}, we list all the Raman-active two-phonon modes for the $D_{4h}$ point group, henceforth expected in \CCOC, according to Bilbao crystallographic server calculations \cite{Bilbao_server}.

In Table~\ref{two_phonon_Raman}, we then list all possible combinations for the two-phonon Raman-active modes for the \CCOC\ case. The calculations are based on the measured phonon energies, except for E$_u$(1) and E$_g$(1), which lie below the measured range, for which we used values calculated from DFT.

\begin{table}[!htb]
    \centering
    \begin{tabular}{c|ccc} 
    \hline \hline 
    symmetry &  \multicolumn{3}{c}{two-phonon} \\ \hline
    $A_{1g}$ & $A_{1g} \otimes A_{1g}$ & $A_{2u} \otimes A_{2u}$ & $B_{u} \otimes B_{u}$ \\
    $E_{g}$ & $E_{u} \otimes A_{2u}$ &  $E_{g} \otimes A_{1g}$ & \\
    $A_{1g}+B_{1g}+B_{2g}$ & $E_{u} \otimes E_{u}$   &  $E_{g} \otimes E_{g}$  & \\
    \hline \hline
    \end{tabular}
    \caption{Raman active two-phonon modes in \CCOC, according to Bilbao crystallographic server calculations \cite{Bilbao_server}.}
    \label{Raman_active_2}
\end{table}

\begin{table*}[!htb]
    \centering
    \begin{tabular}{cc|cc|cc}
    \hline \hline
        Mode & Energy (meV) & Mode & Energy (meV) & Mode & Energy (meV) \\ \hline
        $A_{1g}(1) \otimes A_{1g}(1)$ & 51.94  & $E_{u}(1) \otimes E_{2u}(1)$ & 28.59   &  $E_{u}(2) \otimes A_{2u}(1)$ & 47.72 \\
        $A_{1g}(2) \otimes A_{1g}(2)$ & 64.35  & $E_{u}(2) \otimes E_{2u}(2)$ & 54.1    &  $E_{u}(2) \otimes A_{2u}(2)$ & 49.8  \\
        $A_{1g}(1) \otimes A_{1g}(2)$ & 76.76  & $E_{u}(3) \otimes E_{2u}(3)$ & 88.56   &  $E_{u}(2) \otimes A_{2u}(3)$ & 78.29 \\
                                      &        & $E_{u}(4) \otimes E_{2u}(4)$ & 152     &                               &       \\
        $A_{2u}(1) \otimes A_{2u}(1)$ & 41.44  &                              &         &  $E_{u}(3) \otimes A_{2u}(1)$ & 65    \\
        $A_{2u}(2) \otimes A_{2u}(2)$ & 45.59  & $E_{u}(1) \otimes E_{2u}(2)$ & 41.31   &  $E_{u}(3) \otimes A_{2u}(2)$ & 67.67 \\
        $A_{2u}(3) \otimes A_{2u}(3)$ & 102.58 & $E_{u}(1) \otimes E_{2u}(3)$ & 58.54   &  $E_{u}(3) \otimes A_{2u}(3)$ & 95.57 \\
        $A_{2u}(1) \otimes A_{2u}(2)$ & 43.51  & $E_{u}(1) \otimes E_{2u}(4)$ & 90.26   &                                &      \\
        $A_{2u}(1) \otimes A_{2u}(3)$ & 72.01  &                              &         &  $E_{u}(4) \otimes A_{2u}(1)$ & 96.72 \\
        $A_{2u}(2) \otimes A_{2u}(3)$ & 74.04  &  $E_{u}(2) \otimes E_{2u}(3)$ & 71.28  &  $E_{u}(4) \otimes A_{2u}(2)$ & 98.72 \\
                                      &        &  $E_{u}(2) \otimes E_{2u}(4)$ & 103    &  $E_{u}(4) \otimes A_{2u}(3)$ & 127.2 \\
        $B_{u} \otimes B_{u}$         & 62.44  &  $E_{u}(3) \otimes E_{2u}(4)$ & 120    &                                 &     \\
                                      &        &                               &        &  $E_{g}(1) \otimes A_{1g}(1)$ & 42.68 \\
        $E_{g}(1) \otimes E_{g}(1)$   & 33     &  $E_{u}(1) \otimes A_{2u}(1)$ & 34.98  &  $E_{g}(1) \otimes A_{1g}(2)$ & 55.09 \\
        $E_{g}(2) \otimes E_{g}(2)$   & 46.38  &  $E_{u}(1) \otimes A_{2u}(2)$ & 37.05  &  $E_{g}(2) \otimes A_{1g}(1)$ & 51.77   \\
        $E_{g}(1) \otimes E_{g}(2)$   & 39.68  &  $E_{u}(1) \otimes A_{2u}(3)$ & 65.55  &  $E_{g}(2) \otimes A_{1g}(2)$ & 63.15 \\
        \hline \hline
    \end{tabular}
    \caption{All possible combinations for the two-phonon Raman active mode for the \CCOC\ case.}
    \label{two_phonon_Raman}
\end{table*}

\section{Spectrometer Grating Response Correction}\label{WL_correction}
\begin{figure}[!htb]
   \centering
   \includegraphics[width=0.45 \textwidth]{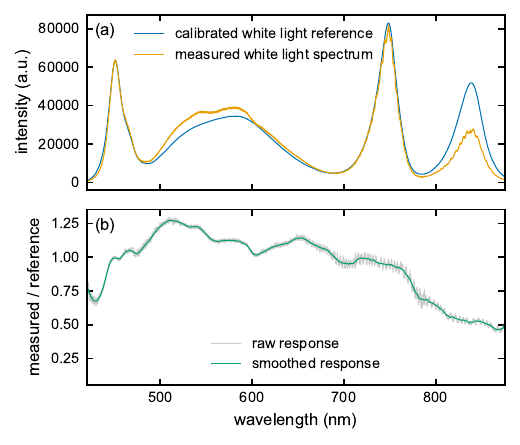}
   \caption{Wave-dependent spectral response correction. (a) Calibrated white-light reference (blue line) and the corresponding measured spectrum (orange line). (b) Instrumental response obtained from the ratio of measured to reference light (grey line), and the smoothed response (green line).}
   \label{fig:WL_correction}
\end{figure}

To correct for the spectrometer grating's wavelength-dependent response, we used a calibrated white-light source. The measured spectrum of this white light and its reference are shown in Fig.~\ref{fig:WL_correction}(a). 
Although the recorded Raman data, the measured white-light spectrum, and the calibrated white-light reference extend beyond 875 nm, the spectrometer's wavelength calibration became unreliable, resulting in a mismatch between the measured and reference spectra. The correction was therefore restricted to wavelengths below 875 nm. 

The response was determined from the ratio of measured white light to its reference spectrum, $R(\lambda) = I_{measured-WL} / I_{reference-WL}$.\\
The resulting response was smoothed using a third-order Savitzky-Golay filter to remove measurement noise and interference fringes originating from CCD etaloning, as shown in Fig.~\ref{fig:WL_correction}(b). The Raman spectra were subsequently corrected according to $ I_{corrected} = I_{measured} / R(\lambda)$.

This correction was applied only to data used for comparing Raman signals acquired with different excitation wavelengths.

\section{Supplementary Figures}\label{Raman_supplemetary}

\begin{figure*}[!htb]
   \centering
   \includegraphics[width=0.9 \textwidth]{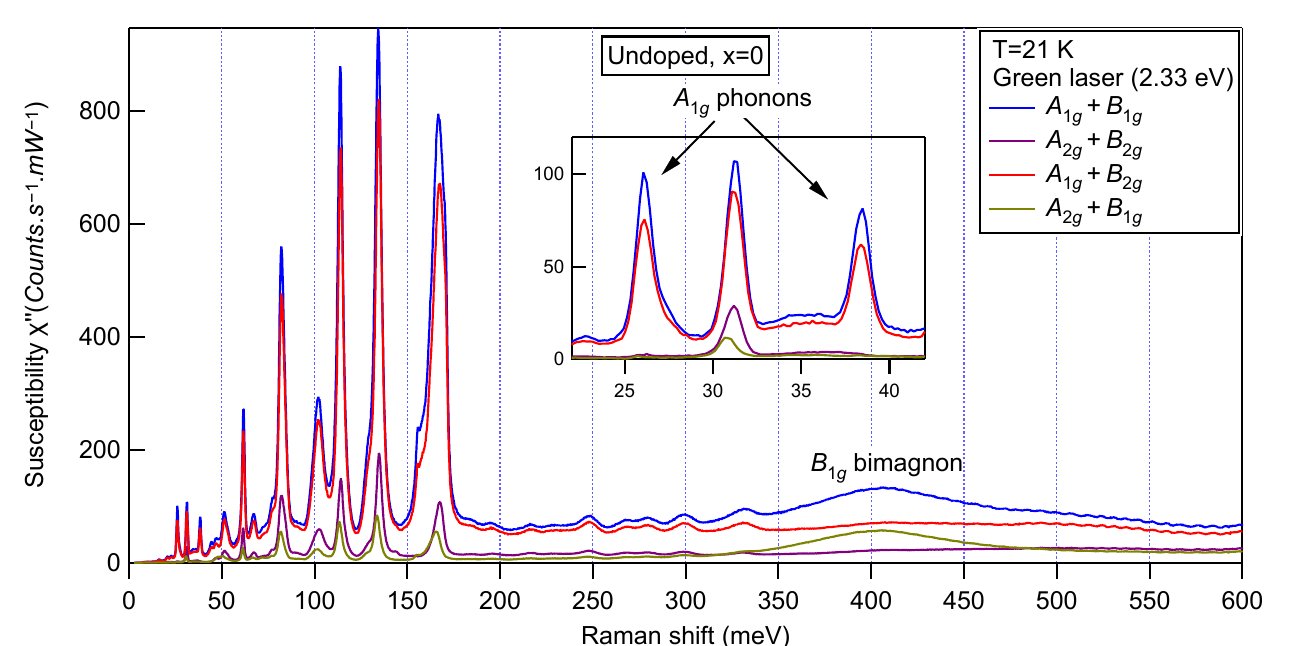}
   \caption{Raman spectra using the four polarisation configurations, taken in the second experiment using a green laser at T~$=21$~K. The inset shows two Raman-active A1g phonons at $\sim26$ and $\sim38$ meV and a Raman-forbidden mode at $\sim30$ meV (see text for details).}
   \label{Green_4config}
\end{figure*}

Figure~\ref{Green_4config} shows a polarization analysis of the undoped sample using a green laser ($E_{incident}=2.33$ eV), with four linear polarization configurations for comparison with the full polarization analysis performed during the preliminary study. Note, however, that here the scattering was at 30$^{\circ}$, and not in back-scattering as in the preliminary test, so a component of the field was along $c^*$, and no circular polarization was measured.

\begin{figure*}[!htb]
   \centering
   \includegraphics[width=0.9 \textwidth]{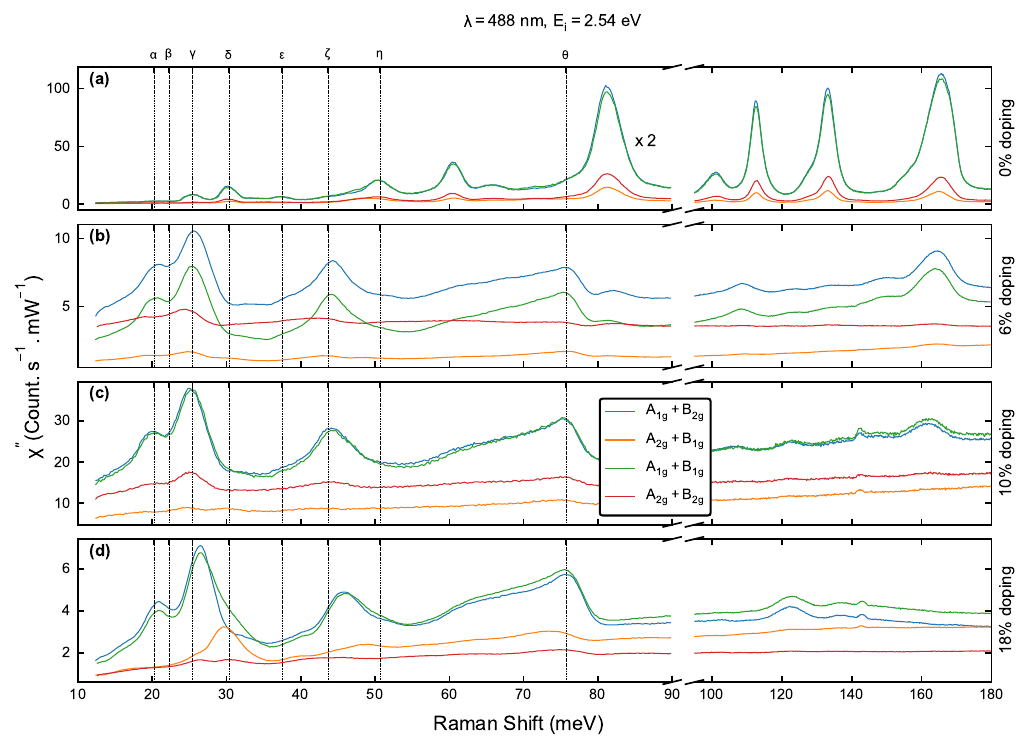}
   \caption{Phonon symmetry analysis for all measured doping levels at 2 K, acquired with the blue laser in all four linear-polarization configurations. The first, second, third, and fourth rows correspond to Na-doping levels of $x = 0,\, 0.6,\, 0.10,\, 0.18$, respectively. The diagonal marks on the x-axis indicate an axis break located around $\omega_D$, marking the upper limit of the one-phonon spectrum. In all panels, the spectra to the left of the axis break were multiplied by 2 for both symmetries. Dashed lines denote the Raman-active modes ($A_{1g}$ labeled as $\gamma$ and $\epsilon$) and the one-phonon Raman-forbidden modes, determined from the undoped compound.}
   \label{phonon_blue_all}
\end{figure*}

\begin{figure*}[!htb]
   \centering
   \includegraphics[width=0.9 \textwidth]{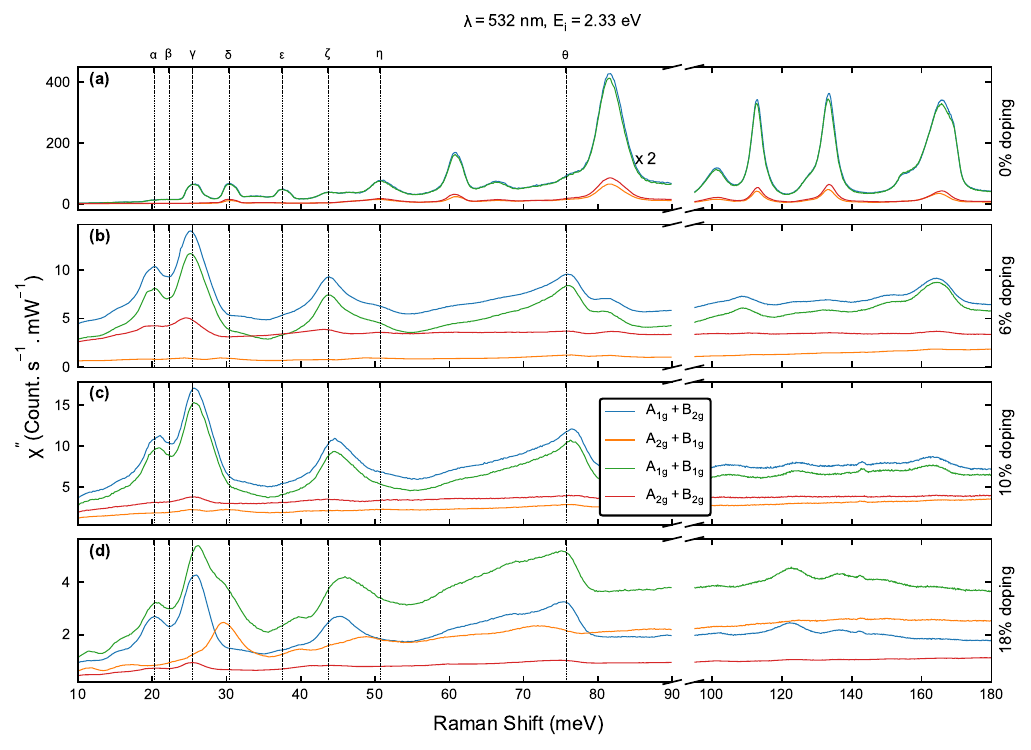}
   \caption{Phonon symmetry analysis for all measured doping levels at 2 K, acquired with the green laser in all four linear-polarization configurations. The first, second, third, and fourth rows correspond to Na-doping levels of $x = 0,\, 0.6,\, 0.10,\, 0.18$, respectively. The diagonal marks on the x-axis indicate an axis break located around $\omega_D$, marking the upper limit of the one-phonon spectrum. In all panels, the spectra to the left of the axis break were multiplied by 2 for both symmetries. Dashed lines denote the Raman-active modes ($A_{1g}$ labeled as $\gamma$ and $\epsilon$) and the one-phonon Raman-forbidden modes, determined from for the undoped compound.}
   \label{phonon_green_all}
\end{figure*}

\begin{figure*}[!htb]
   \centering
   \includegraphics[width=0.9 \textwidth]{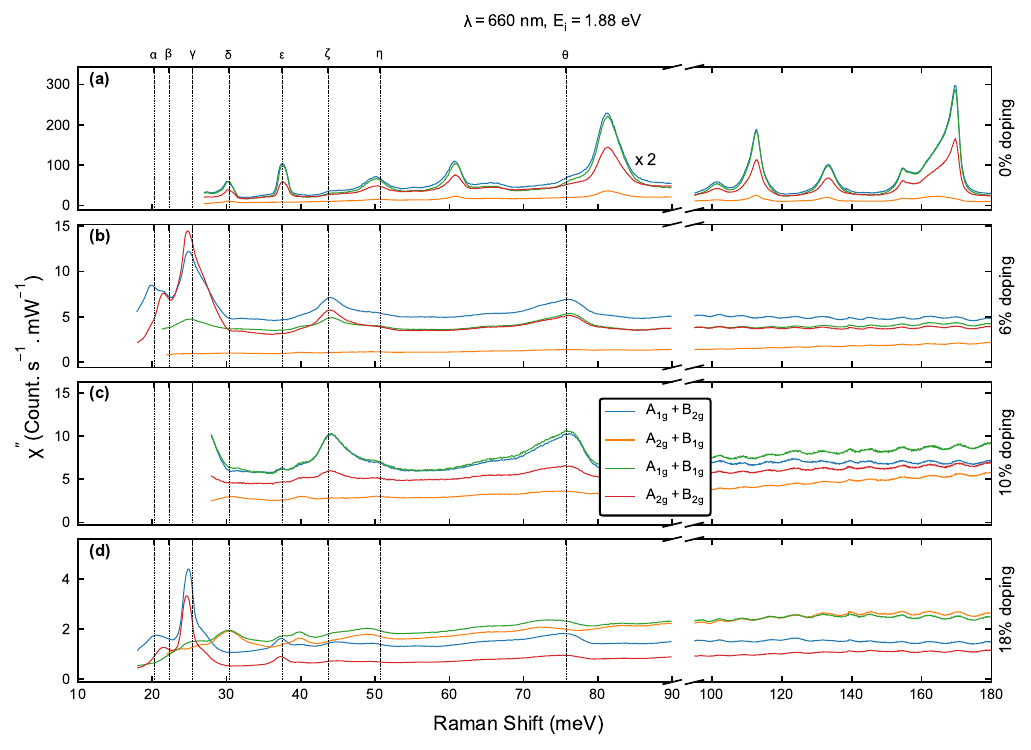}
   \caption{Phonon symmetry analysis for all measured doping levels at 2 K, acquired with the red laser in all four linear-polarization configurations. The first, second, third, and fourth rows correspond to Na-doping levels of $x = 0,\, 0.6,\, 0.10,\, 0.18$, respectively. The diagonal marks on the x-axis indicate an axis break located around $\omega_D$, marking the upper limit of the one-phonon spectrum. In all panels, the spectra to the left of the axis break were multiplied by 2 for both symmetries. Dashed lines denote the Raman-active modes ($A_{1g}$ labeled as $\gamma$ and $\epsilon$) and the one-phonon Raman-forbidden modes, determined from the undoped compound.}
   \label{phonon_red_all}
\end{figure*}

Figures~\ref {phonon_blue_all}, \ref{phonon_green_all}, and \ref{phonon_red_all} show the complete set of all four linear-polarization configurations for all measured doping levels using blue, green, and red lasers, respectively.

\section{Fluorescence}
\label{fluo}
\begin{figure*}[!htb]
   \centering
   \includegraphics[width=1 \textwidth]{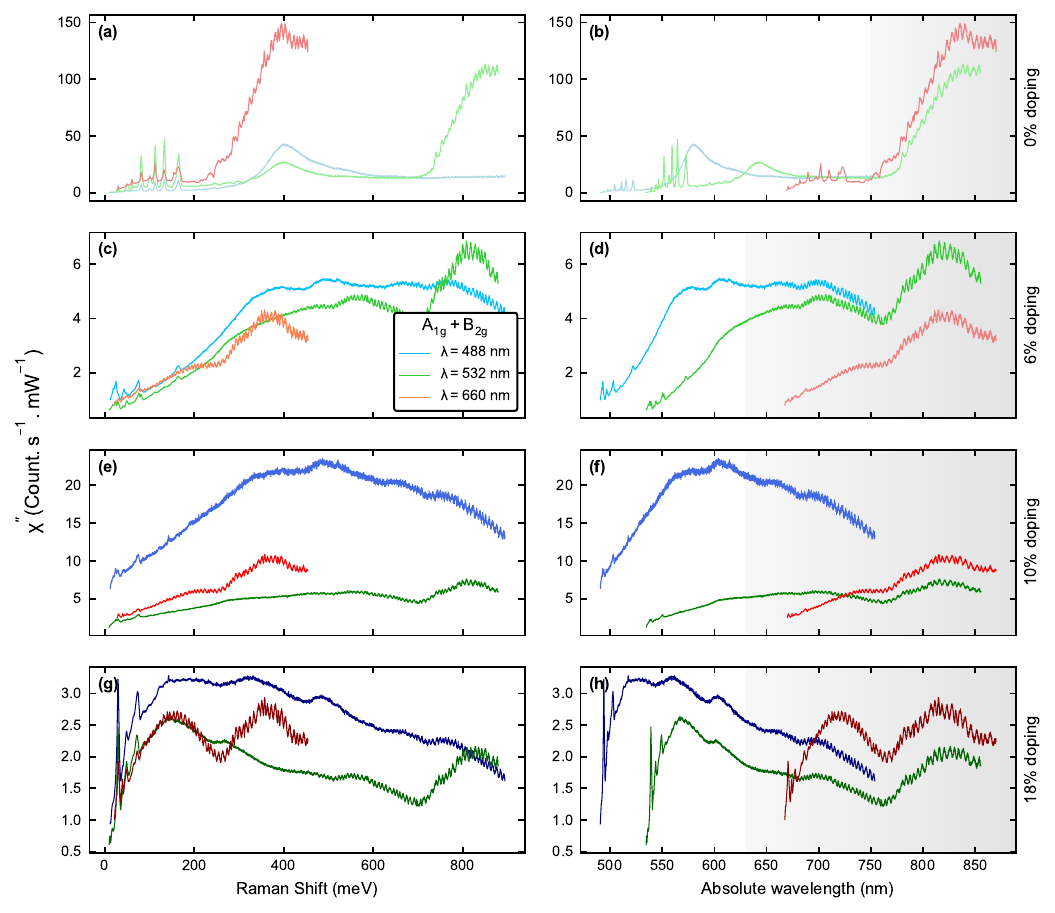}
   \caption{Comparison of ``$A_{2g} + B_{1g}$" Raman signal using all three lasers, Blue, green, and red, plotted \textit{vs} the Raman shift in the left column and \textit{vs} the absolute wavelength nm in the right column. The first, second, third, and fourth rows correspond to Na-doping levels of $x = 0,\, 0.6,\, 0.10,\, 0.18$ respectively.
   The shaded region highlights the fluorescence contribution.}
   \label{fig:fluo}
\end{figure*}

To identify the fluorescence signal in our Raman spectra, we plotted spectra measured with all three lasers (blue, green, and red) \textit{vs} the Raman shift (cm$^{-1}$) and the absolute energy (nm) in Fig.~\ref{fig:fluo}. The fluorescence contribution is distinguished based on the peaks that remain at fixed absolute energy (nm) upon detuning the excitation energy, unlike the behavior of phonons and magnetic excitations, whose energies shift accordingly.
Fig.~\ref{fig:fluo}(a) shows the ``$A_{2g} + B_{1g}$" symmetry plotted \textit{vs} the Raman shift(cm$^{-1}$), in the undoped \CCOC, where the bimagnon peak is at 400 meV in all three spectra measured with blue, green, and red lasers. However, in Fig.~\ref{fig:fluo}(b), we show that the peak that appears at 400 meV using the red laser is dominated by the fluorescence signal, which is highlighted by the shaded region.\\
The fluorescence shifts towards lower energy upon doping, as highlighted in Fig.~\ref{fig:fluo}(d, f, and h).
We note that in the maximally doped sample ($n=0.18$), even for the red light, when the bimagnon peak lies within the fluorescence region, it still dominates the spectral response.

\section{Temperature dependence in the undoped sample}

\begin{figure*}
	\includegraphics[width=1 \linewidth]{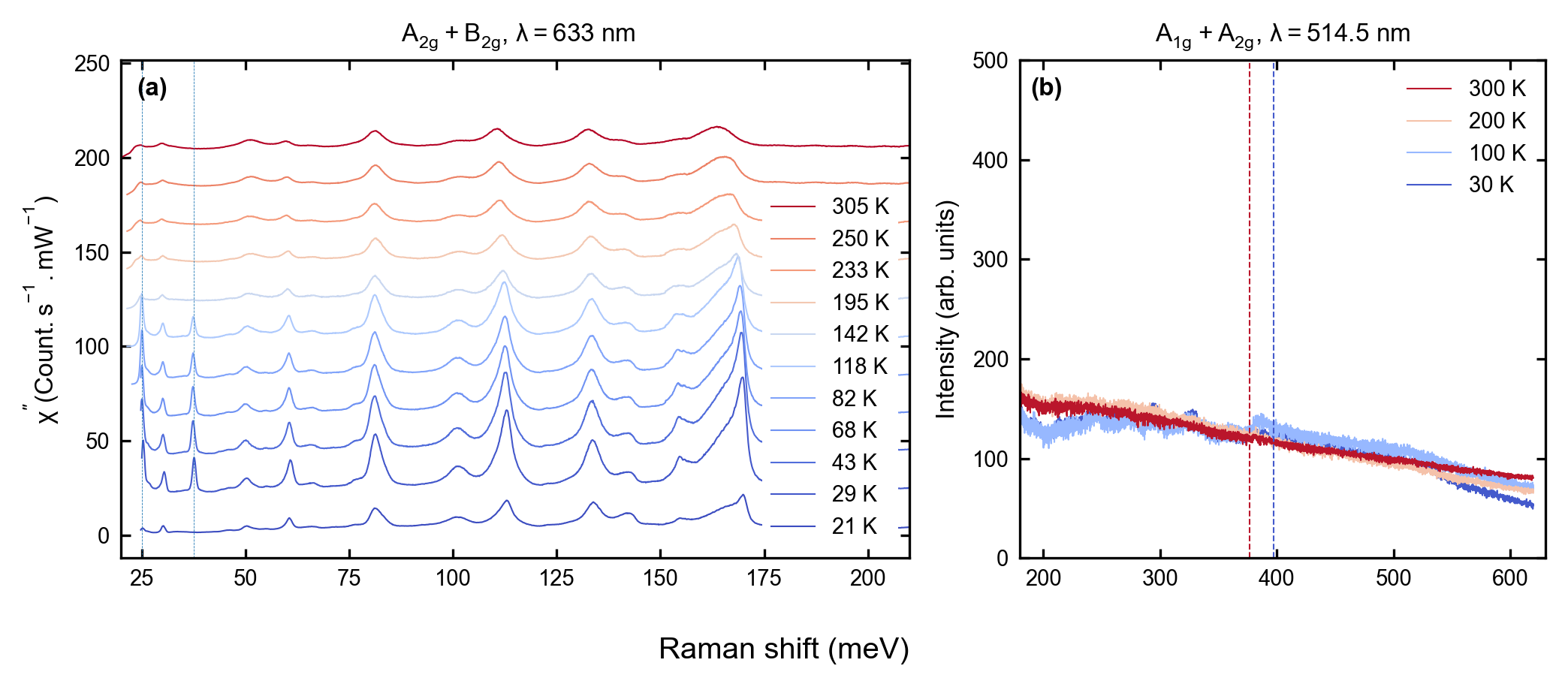}
	  \caption{\label{RamanC_TdepA2g}
      (a) Raman scattering of the A$_{2g}$+B$_{2g}$ symmetry from linearly polarized light, from 21 K up to room temperature, using a red laser for improved energy resolution. Two dashed lines mark the position of the A$_{1g}$. 
      (b) Raman scattering of the A$_{1g}$+A$_{2g}$ symmetry from circularly polarized light, from 30 K up to room temperature using a green laser. Red and blue lines mark the bimagnon energy at room temperature and 30~K, respectively.}
 \end{figure*}

In Figure~\ref{RamanC_TdepA2g}(a), we show a detail of the Raman scattering in the undoped, parent compound \CCOC\, with A$_{2g}$+B$_{2g}$ symmetry from linearly polarized light, from 21 K up to room temperature, using a red laser for improved energy resolution. Two dashed lines mark the position of the A$_{1g}$, which should not be observed in this configuration. However, a relatively strong signal emerges between approximately 200 K and 20 K, the origin of which is unclear. Further investigations are underway using neutron and X-ray diffraction to ascertain whether there is a crystal phase transition to a structure different from the known $I4/mmm$. 

In Figure~\ref{RamanC_TdepA2g}(b), we show, for the same antiferromagnetic parent compound \CCOC\, the Raman scattering of the A$_{1g}$+A$_{2g}$ symmetry from circularly polarized light, from 30 K up to room temperature using the green laser of the ``Paris-setup", showing that the high-energy signal assigned to bimagnon scattering stays in the B$_{1g}$ symmetry at all temperatures. 

\section{Bimagnon energy determination}
\label{Bimagnon_energy_det}
\begin{figure*}[!htb]
   \centering
   \includegraphics[width=1 \textwidth]{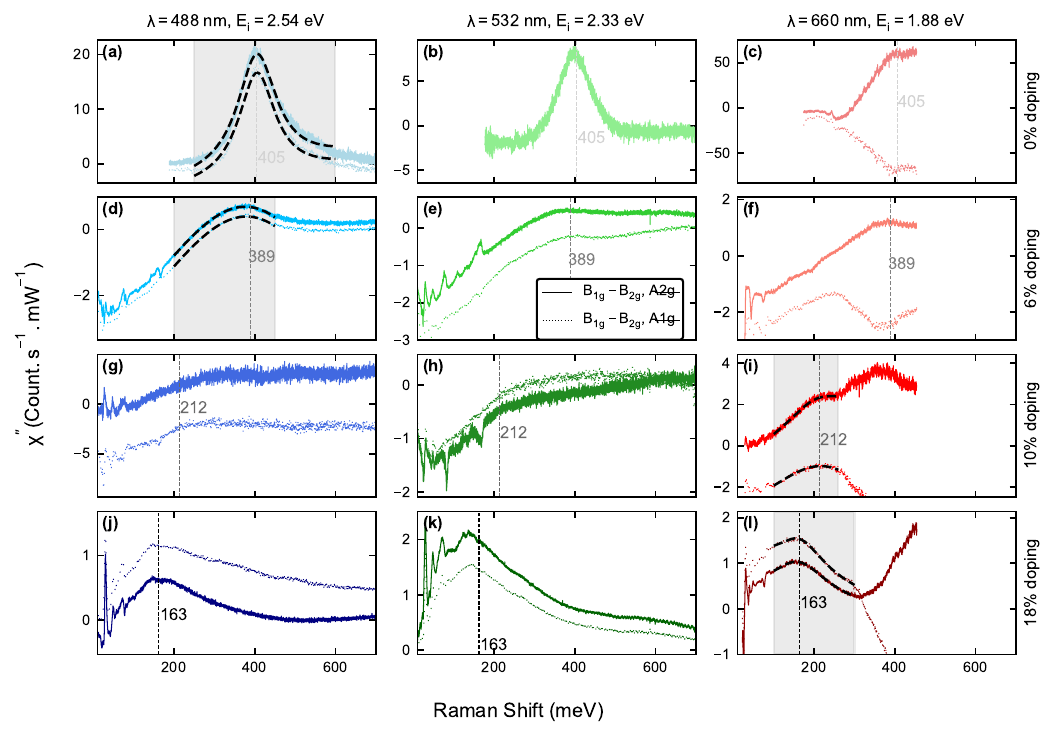}
   \caption{Laser and doping dependence of the bimagnon at 2 K shown in the $B_{1g} - B_{2g}$ channel. The first, second, and third columns correspond to the blue, green, and red lasers, respectively.  The first, second, third, and fourth rows correspond to Na-doping levels of $x = 0,\, 0.6,\, 0.10,\, 0.18$, respectively.
   The solid lines represent the ``$B_{1g} - B_{2g}$, \sout{$A_{2g}$}", obtained by substracting the $A_{2g} + B_{2g}$ from $A_{2g} + B_{1g}$ signal. The dotted line represents the ``$B_{1g} - B_{2g}$, \sout{$A_{1g}$}", obtained by substracting the $A_{1g} + B_{2g}$ from $A_{1g} + B_{1g}$ signal.
   Vertical dashed lines show the bimagnon positions, determined from fits to selected spectra (dashed lines) in the shaded regions.}
   \label{B1gmB2g}
\end{figure*}

In order to determine the bimagnon energy position in the main experiment (Grenoble set-up, with only linear polarization), we subtracted the $A_{2g} + B_{2g}$ from $A_{2g} + B_{1g}$, and $A_{1g} + B_{2g}$ from $A_{1g} + B_{1g}$ signals, to obtain the $B_{1g} - B_{2g}$ symmetry, denoted as "$B_{1g} - B_{2g}$, \sout{$A_{2g}$}" and "$B_{1g} - B_{2g}$, \sout{$A_{1g}$}" respectively as shown in Fig~\ref{B1gmB2g}. This better isolates the $B_{1g}$ symmetry, where the bimagnon appears. These two resulting $B_{1g} - B_{2g}$ signals were used to fit the bimagnon peak with a Lorentzian function and a linear background. The final energy position was taken as the average of the energy values obtained from the two fits.

For the undoped \CCOC\ and $n=0.06$ doping, the enhancement is strongest with the blue laser; therefore, we fitted the corresponding spectra. For the $n=0.10$ and $n=0.18$ doping, the enhancement is strongest in the red laser; therefore, the corresponding spectra were used for the fit.

We note that the negative dip in our signal in Fig~\ref{B1gmB2g}(c,f,i,l) is likely caused by the dependence of the fluorescence signal on the incident polarisation. Since the $A_{1g} + B_{2g}$ and the $A_{1g} + B_{1g}$ geometries differ by a $45^{\circ}$ rotation of both the incident and scattered polarizations, the fluorescence background can exhibit a different intensity in the two measurements (see section~\ref{fluo}).

\end{document}